\documentclass[manuscript]{aastex}   
\usepackage{emulateapj5}             
\usepackage{natbib}
\shorttitle{The Structure of Stellar Coronae in Active Binary Systems}
\shortauthors{Sanz-Forcada, Brickhouse \& Dupree}

\newcommand{\gl}{$\lambda$}

\begin{document}

\title{The Structure of Stellar Coronae in Active Binary Systems}
\author{J. Sanz-Forcada\altaffilmark{1,2},
  N. S. Brickhouse\altaffilmark{1}, and A. K. Dupree\altaffilmark{1}}
\altaffiltext{1}{Harvard-Smithsonian Center for Astrophysics; 60 Garden St.,
Cambridge, MA 02138 (USA)}
\altaffiltext{2}{INAF -- Osservatorio Astronomico di Palermo;
  Piazza del Parlamento, 1; Palermo, I-90134 (Italy)}
\email{jsanz@astropa.unipa.it, nbrickhouse@cfa.harvard.edu,
  adupree@cfa.harvard.edu}

\begin{abstract}

A survey of 28 stars (22 active binary systems, plus 6 single stars or
wide binaries for comparison) using extreme ultraviolet spectra has been conducted to
establish the structure of stellar coronae in active binary systems
from the emission measure distribution (EMD), electron densities, and
scale sizes. Observations obtained by the {\it Extreme Ultraviolet
Explorer} satellite (EUVE) during 9 years of operation are included
for the stars in the sample.  EUVE data allow a continuous EMD to be
constructed in the range log~T$_e(K)\sim$5.6--7.4, using iron emission
lines. These data are complemented with IUE observations to model
the lower temperature range (log~T$_e[K]\sim$4.0--5.6).
Inspection of the EMD shows an outstanding narrow enhancement, or
``bump'' peaking around log~T$_e(K)\sim$6.9 in  25 of the stars,
defining a fundamental coronal structure.  The emission measure per
unit stellar area decreases with increasing orbital (or photometric)
periods of the target stars; stars in binaries generally have more material at coronal
temperatures than slowly rotating single stars.  High electron densities ($N_e\ga
10^{12}~cm^{-3}$) are derived at $\sim$log~T$_e(K)\sim$7.0 for some targets, implying
small emitting volumes.

The observations suggest the magnetic stellar coronae of these stars
are consistent with two basic classes of magnetic loops: solar-like
loops with maximum temperature around log~T$_e(K)\sim$6.3 and lower
electron densities ($N_e\ga 10^{9}-10^{10.5}~cm^{-3}$), and hotter
loops peaking around log~T$_e(K)\sim$6.9 with higher electron
densities ($N_e\ga 10^{12}~cm^{-3}$).  For the most active stars,
material exists at much higher temperatures (log~T$_e[K]\ge$6.9) as well.
However, current {\it ab initio}
stellar loop models cannot reproduce such a configuration.  Analysis
of the light curves of these systems reveals signatures of rotation of
coronal material, as well as apparent seasonal (i.e. year-to-year) changes in the activity
levels.

\end{abstract}


\section{Introduction}
The study of coronal structure from early X-ray and EUV satellites has
generally been limited to 2 or 3 temperature emission measure
fits. After the launch of the {\it Extreme Ultraviolet Explorer}
satellite (EUVE) a continuous emission measure distribution (EMD) in
the coronal region has been obtained for a few objects, but no
systematic study has been carried out to date in a substantial set of
stars.  Early EUVE observations have shown a quite different coronal
structure in active stars from that of the solar corona
\citep{dup93}. After nine years of EUVE data collection, many cool
stars have been observed \citep[see, for instance,][]{crai97}, some of
them several times, allowing the 
acquisition of good spectra for many stars so that reliable EMDs can
be calculated.  A survey of 28 stars has been conducted \citep{tesis}
to find stellar parameters that can be related to the observed coronal
emission. In this study a total of 22 active binary systems (in
particular RS~CVn and BY~Dra systems), and 6 single stars or wide
binaries has been included.  This sample covers a wide range of
luminosity class, spectral type, and rotational period (see
Table~\ref{tabparam}), and hence differences might be expected to
occur in their coronae that can be related to stellar parameters.  In
this paper we present the results for 21 stars, complementing our
previous analysis \citep*{paper1}, which includes the stars V711~Tau,
UX~Ari, $\sigma$~Gem, II~Peg, $\beta$~Cet (also studied here with
recent observations), and AB~Dor. Two other binaries, Capella and
$\lambda$~And 
(Dupree et~al. 1993; Dupree, Brickhouse, \& Sanz-Forcada 2002; 
{Sanz-Forcada}, {Brickhouse}, \& {Dupree} 2002)
complete the set of 28 stars.

The observed EMD of Capella \citep{dup93} revealed the presence of a
narrow enhancement or ``bump'' at log~T(K)$\sim$6.8, that varies
little in observations taken at different epochs \citep{dup02}. Other
stars show similar structure including $\lambda$~And \citep{sanz01},
and the six stars in \citet{paper1}, some of which are also studied by
\citet{gri98}.  Analysis of the changes observed in the EMD during
large flares has also been carried out for 6 stars: $\lambda$~And
\citep{sanz01}, V711~Tau, UX~Ari, $\sigma$~Gem, and II~Peg
\citep{paper1}, and AR~Lac (this work), showing that the bump remains
and is stable in temperature, while the emission measure increases
during flares.

In this paper we describe the detailed results for the remaining 21
stars of the sample and the global conclusions from the entire sample.
The stars have been grouped according to their observed coronal
spectra: a first group of low and intermediate activity stars,
dominated by lines formed at log~T(K)$\sim$5.8--6.5, with different
levels of flux in lines formed at higher temperatures; a second group
of ``active'' stars, with spectra dominated by lines formed at
log~T(K)$\sim$6.7--7.1; and a third group with a very significant
presence of even hotter material (indicated by Fe XXIII-XXIV).

In \S~\ref{sec:observations} we discuss the EUVE and IUE observations,
followed by the techniques employed in the analysis of the data
(\S~\ref{sec:analysis}). Individual results, following the
classification of the three groups, are described in
\S~\ref{sec:results}. A general discussion of these results and a
comparison between the different degrees of activity are made in
\S~\ref{sec:discussion}, and the Conclusions are summarized in
\S~\ref{sec:conclusion}.

\section{Observations}\label{sec:observations}

EUVE observations taken between 1993 January and 2000 September are
used (Table~\ref{euvetimes}).  Some of the observations were awarded
to us through the Guest Observer program, while most of them were made
available through the Multimission Archive at Space Telescope (MAST).
EUVE spectrographs cover the spectral range 70--180~\AA, 170--370~\AA, 
and 300--750~\AA\ for the short-wavelength (SW), medium-wavelength
(MW), and long-wavelength (LW) spectrometers respectively, with
corresponding spectral dispersion of $\Delta\lambda\sim$ 0.067, 0.135,
and 0.270~\AA/pixel, and an effective spectral resolution of
$\lambda/\Delta\lambda$$\sim$200--400. The Deep (DS) Survey Imager has
a band pass of 80--180~\AA\ \citep*{hai93}.

\begin{figure}\
\epsscale{.80}
\plotone{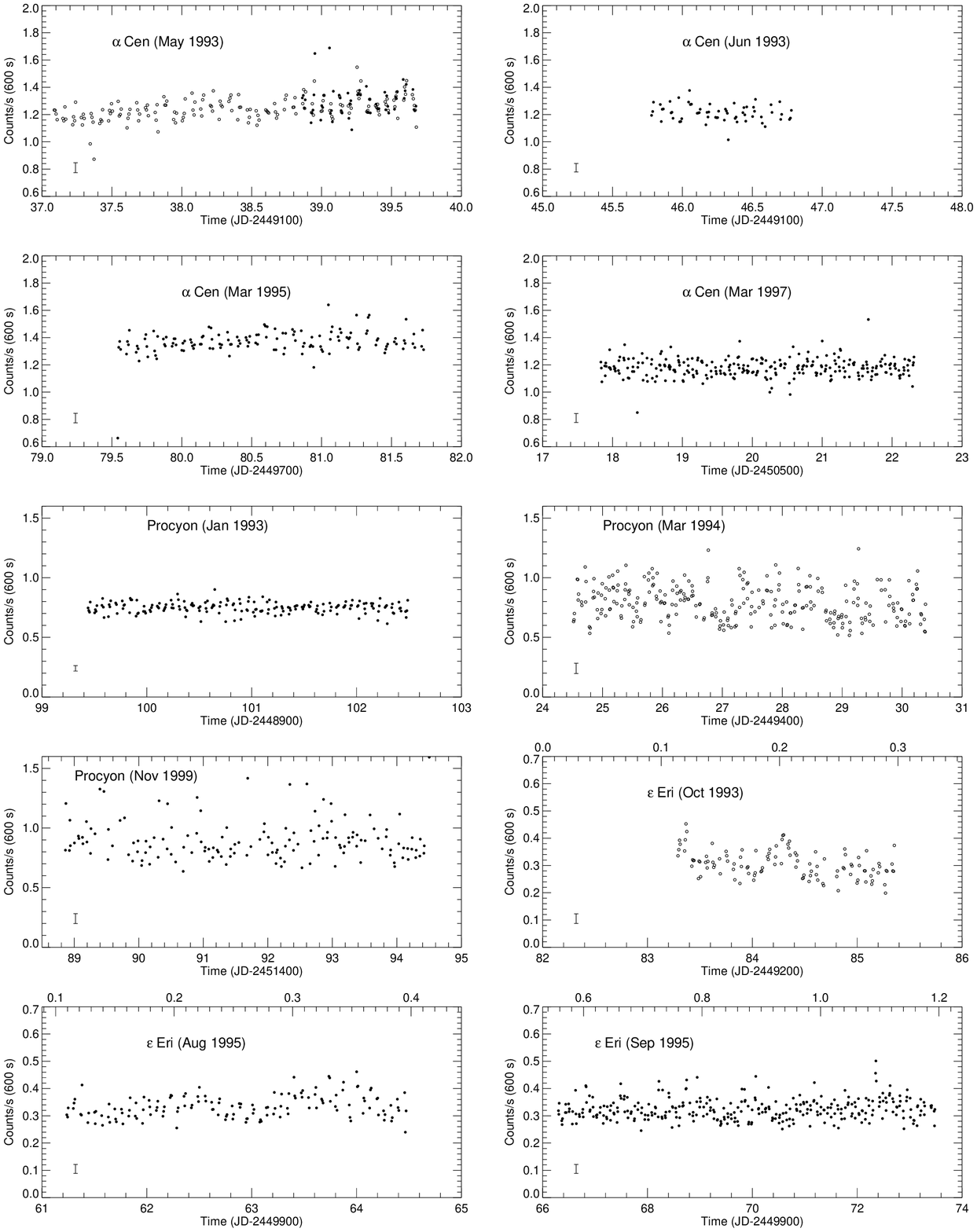}
  \caption{DS light curves as a function of Julian Date (lower axis)
  and orbital phase (upper axis). We use the convention that at
  orbital phase 1.0 the primary  star is located behind the secondary
  star (see Table~\ref{tabparam}); photometric phase is used in the
  upper axis for $\epsilon$~Eri and LQ~Hya, and no periods are
  available for
  $\alpha$~Cen, Procyon, and $\beta$~Cet. Open circles mark data affected
  by the dead spot (and corrected by the DS to SW flux ratio,
  see \S~\ref{sec:observations}), while solid circles represent unaffected
  data. An average one-sigma error bar is shown as
  reference on the left side of each plot. Only points with S/N
  higher than 5 are plotted. The bin size is 600~s. 
  \label{plc}}
\end{figure}

EUVE light curves (Fig.~\ref{plc}) were built from the DS image, by
taking a circle centered on the source, and subtracting the sky
background within an annulus around the center.  Standard procedures
were used in the IRAF package EUV v.1.9.  Time bins are 600~s. Points
affected by the ``dead spot'' are marked as open circles in the light
curves, while filled circles mark the corrected points. The ``dead
spot'' is a low gain area of the DS detector that affects some of the
observations taken in 1993 and 1994, resulting in variable levels of
contamination of the signal \citep[see][]{mill95}. We corrected
the effects of the dead spot contamination by ratioing unaffected DS
flux to the flux from the integrated SW spectrum measured
simultaneously.  This 
gives a correction factor that can be applied to the
DS points affected by the dead spot.  When all the DS flux
points are contaminated, we normalized using a star 
with unaffected DS flux, and a similar emission 
measure distribution. The error bar 
included in the figures indicates the average of the non-contaminated
DS fluxes.  Spectra for each star, extracted
from each spectrograph, are binned over the total observation 
and then summed for the whole set of
observations. Fig.~\ref{specs} shows the SW and MW spectra for all the 
targets, and Fig.~\ref{specesp} contains the LW spectra for 5 of the 
stars.  The addition
of spectra from different observations is not of concern since no
significant degradation was reported in the performance of the EUVE
spectrographs during the mission \citep{abb96}.\footnote{Results of
the final EUVE calibration observations made during the last month of
EUVE science operations in January 2001 demonstrated: (1) no
significant degradation in the SW spectrometer; (2) possibly up to
15\% loss of sensitivity non-uniformly in the MW spectrometer; (3) no
change (to $<$10\%) in the LW detector when comparison of 1993 and
2001 spectra are made.  The LW detector showed some degradation in
1999, but the detector recovered by the end of the mission.  The DS
count rates are in agreement within 10\% of previous
measurements. (See {\it
ftp://legacy.gsfc.nasa.gov/euve/doc/final\_calib.report}.)}  Lines
identified in the summed EUVE spectra are given in
Table~\ref{euveflux} and Table 4.  Our primary goal is to identify Fe diagnostic
lines, but we also include strong lines from other elements.

Spectra from the {\it International Ultraviolet Explorer} (IUE)
archive (NEWSIPS extractions) have also been used to construct the EMD
curve of the stars by providing lines formed at lower temperatures
than those occurring in the EUVE spectra. Low resolution spectra
($\sim$6 \AA) covering $\lambda\lambda$~1100--1950 were employed. In
the case of AR~Lac, quiescent and active spectra have been selected,
with discrimination based on the changes observed in line fluxes found
in different spectra. IUE line fluxes used to determine the EMD are
listed in Table~\ref{tabiue}.

\section{Data Analysis}\label{sec:analysis}

The DS light curves of the targets are shown in Fig.\ref{plc} where 
they are compared with the orbital phase for the binaries.  For
$\epsilon$~Eri and LQ Hya, a photometric phase is displayed; no
periods are considered for $\alpha$~Cen, $\beta$~Cet and Procyon.

To obtain fluxes of the individual EUV emission lines we first
performed optimized extractions from the summed two-dimensional images
by removing an averaged background evaluated on either side of the
spectrum, using the software provided in IRAF and the EGODATA 1.17
reference data set.  A local continuum in the spectrum itself,
determined by visual inspection, was subtracted from each line where
necessary.  The error in the line flux is defined as
$\sigma=1/[S+B(1+1/n)]^{1/2}$, where S is the net signal, B is the
estimated average background, and n is the oversampling ratio (i.e.,
the ratio of total background pixels to the number of total source
spectral pixels in the image), having a value n$\sim$10--15 in our
extraction.

To correct the observed fluxes for interstellar hydrogen and helium
continuum absorption, we used a ratio \ion{He}{1}/\ion{H}{1}=0.09
\citep{kimb93}, and values for the hydrogen column density obtained in
different ways for each star.  For some targets, direct measurements
of the column density were available from Lyman series absorption
features. Frequently the observed ratios of the \ion{Fe}{16} \gl335
and \gl361 lines can indicate the amount of interstellar absorption
because the theoretical ratio (1.94 in photon units) is determined
from fundamental atomic physics.  When these line fluxes are
available, they have been used to establish or corroborate hydrogen
column densities to the targets (see Fig.~\ref{ismcalc}). When those
values were not accurate enough, we deduced the column density from
tabulations \citep{fru94} of stars nearby in the sky, and these were
the adopted values if no additional references are given.
Table~\ref{tabparam} lists the values assumed. Further discussion
follows in the sections for the individual stars.

The electron density in the corona of the stars at
log~T$_e(K)\sim6.9-7.0$ has been inferred from ratios of the observed
fluxes (corrected for interstellar absorption) of
\ion{Fe}{19}~$\lambda$91.02/$\lambda$108.37, \ion{Fe}{20}
$\lambda$110.63/($\lambda$118.66+$\lambda$121.83),
\ion{Fe}{21}~$\lambda$102.22/$\lambda$128.73,\\
\ion{Fe}{21}~($\lambda$142.16+$\lambda$142.27)/$\lambda$102.22, and
\ion{Fe}{22} $\lambda$114.41/$\lambda$117.17 in the summed spectra.
Iron line emissivities were generally computed for the densities
derived in each spectrum, when available.  Atomic models for
\ion{Fe}{20}--{\small\sc XXII} were taken from \citet*{bri95}, with
\ion{Fe}{19} from Liedahl's HULLAC calculations
\citep[see][]{bri98}. These models have recently been compared with
measured tokamak spectra at different densities in the range $10^{12}$
-- $10^{14}~cm^{-3}$ at $\sim 10^7$~K and show good agreement
\citep{four01}. Table~\ref{tabne} shows the results for each
star.

We performed a line-based analysis of the emission spectra in order to
calculate the EMD ($\int N_e N_H dV$ cm$^{-3}$, where $N_e$ and $N_H$
are electron and hydrogen densities, in cm$^{-3}$) corresponding to
the observed fluxes.  In contrast to ROSAT and ASCA measurements,
which assume coronal models with only 2 or 3 temperatures, EUVE
gives information on a continuous set of ionization states.  In fact,
all stages of iron ionization are represented from \ion{Fe}{9} through
\ion{Fe}{24} except for \ion{Fe}{17}, which has no strong transitions
in the EUV spectral range.  We used the line emissivities calculated
from \citet{bri95} for the EUVE iron lines, based on a solar iron
abundance\footnote{The solar iron abundance is defined as (12. $+$
log~$Fe \over H$), where $Fe \over H$ represents the ratio of iron to
hydrogen by number.}  of 7.67 \citep{anders}. Line emissivities from
\citet{raym88} are used for the (non-iron) lines formed in the UV
region.  Theoretical fluxes were calculated using the assumed EMDs
\citep[see][ and references therein]{dup93,bri98} which were then
iterated to obtain the EMD that best matches the observed
fluxes. Generally agreement is better than a factor of two. It is
important to note that we have integrated over the entire atomic
emissivity function for each line in order to predict the model
fluxes, and do not simply assume formation at a single temperature or
temperature range.  Iron lines severely contaminated by lines of other
elements and some complex blends have been estimated by using the
Astrophysical Plasma Emission Code (APEC) v1.10 \citep{smith01}.
These are excluded from the EMD analysis as marked in
Table~\ref{euveflux}. Figure~\ref{emdfigs} shows the EMD of summed
spectra of the stars in the sample.  Values used for the EMD are given
in Tables~\ref{tabemd} and \ref{tabemd2}, with a simple
characterization suggested in Table~\ref{slopes}.

\section{Results}\label{sec:results}

\subsection{Low activity levels}\label{sec:lowactive}

In this group we include stars showing a spectrum dominated by lines
formed at\\ log~T(K)$\sim$5.8--6.5 (\ion{Fe}{9}--{\small \sc XVI}),
although with different contributions from the lines formed at higher
temperatures (see Figs.~\ref{specs}a, \ref{specesp}).   In
the EMD derived for these stars, $\alpha$~Cen represents the lowest
activity level observed in the sample, both in the transition 
region and in the corona. Procyon also has low levels 
in the corona, but the transition region EMD is comparable to 
that observed in more active stars (see \S~\ref{sec:discussion}). 
Finally, $\epsilon$~Eri and $\xi$~UMa have high emission measure 
in the two temperature
ranges, such that their EMDs represent an intermediate step towards
the second group included in the sample.

\subsubsection{$\epsilon$~Eri}
$\epsilon$~Eri (HD 22049, HR 1084) is a relatively young star
($\sim$1~Gyr) showing high levels of activity.  The effects of the
dead spot prevent an analysis of the seasonal (i.e. year-to-year)
variations of the light curves, but small scale variations can be seen
with a frequency of $\sim$1--1.5 days, and some flaring activity could
be present in the second part of the 1995 observations
(Fig.~\ref{plc}).

Studies of the EUVE observations have been carried out by
\citet*{lami96} and \citet{sch96}. \citet{lami96} derived the emission
measure of each individual line as if it were emitting only at its
maximum temperature thus obtaining an upper limit to the EMD
value. This model peaks at log~T(K)$\sim$6.5. Densities derived from
\ion{Fe}{14} line ratios \citep{lami96} yield a value of
log~N$_e$(cm$^{-3}$)$\sim$9.5 at log~T(K)$\sim$6.2, similar to results
found by \citet{sch96}.

We added all the available EUVE data of $\epsilon$~Eri in order to
improve the statistics of the spectra, and the accuracy of the EMD.
Although the EMD calculated (Fig.~\ref{emdfigs}) indicates a peak
around log~T(K)$\sim$6.4, a range of values of the EMD 
predicts very similar line
fluxes.  It is difficult to distinguish a peak around
log~T(K)$\sim$6.5, from two peaks at log~T(K)$\sim$6.4 and
log~T(K)$\sim$6.8.  In order to identify a preferred EMD model, it
would be optimum to use lines with a maximum of emission around
log~T(K)$\sim$6.5--6.7 such as Fe XVII which are not available in the
EUVE spectral range.

The electron density calculated from the \ion{Fe}{21} lines
(3$\times$10$^{13}$ cm$^{-3}$, see Table~\ref{tabne}) points towards
the presence of two kinds of structures at log~T(K)$\sim$6.2 and
log~T(K)$\sim$7.0, because the inferred densities differ by $\sim$4
orders of magnitude, although some caution must be exercised in the
interpretation of the measured electron density at high temperatures
(see \S~\ref{sec:discussion}) in this star (the adopted value in the EMD 
was log~N$_e$[cm$^{-3}$]$\sim$13.0).  The observed levels of
EUV/flaring activity are higher than the flux levels of the quiet Sun,
creating different conditions for the atmosphere of the planet found
by \citet{hat00} associated with $\epsilon$~Eri.

\subsubsection{Procyon}

Procyon ($\alpha$~CMi, HD 61421, HR 2943) is an F5~IV-V star
frequently used for comparison with the Sun.  Its rotational period of
9.06~d is estimated from a v$\cdot$sin~i=6.1 $km\ s^{-1}$
\citep{med99}.  The DS light curve of Procyon shows the presence of
short-term modulation in a non-periodic time pattern of $\sim$14~hr in
1993 (Fig.~\ref{plc}).  The observations taken in 1999 show higher
dispersion in flux values than in 1993 and 1994; however, this effect can be due to the
increase of activity in the Sun, as has been found in other
observations taken in 1999 and 2000. In any case it does not seem
possible to attribute these variations to intrinsic changes in the
star.

During the EMD fitting process in Procyon, some of the lines have
shown larger discrepancies between the observed fluxes and those
predicted by the atomic model of \citet{bri95}. In particular, the
\ion{Fe}{12} $\lambda$364.4 line seems too strong, as if blended,
though the APEC model does not show an obvious candidate for
blending.

\subsubsection{$\xi$~UMa}
$\xi$ UMa is a multiple system formed by four stars grouped in two
spectroscopic binaries \citep{griffin98} in a visual orbit of
$\sim$60~yr.  EUVE can not resolve the components.  $\xi$~UMa A (HD
98231) is a spectroscopic binary with a P$_{orb}$=670.24 days
\citep{griffin98}.  A G0V star is the only star observed, and the
companion is unknown -- perhaps an M star \citep{griffin98}.
$\xi$~UMa Aa shows low activity levels in the chromosphere, having
very weak emission in the \ion{Ca}{2} H \& K lines \citep{mon95}.
$\xi$~UMa~B (HD 98230) shows an active spectrum, with stronger
\ion{Ca}{2} H \& K emission lines coming from the observed G5V star,
and a possible contribution from an unseen late-K dwarf
\citep{griffin98}.  The low value of the mass function
(f[m]=0.000046M$_{\odot}$) found by \citet{griffin98} points towards a
very small value of the inclination ($\la 11\arcdeg$), meaning a
system observed almost at the pole. From the effective temperature
\citep[T$_{eff}$=5650~K,][]{cayr94}, and the typical values of stellar
radii available in \citet{gray92} relating radii and mass with
spectral type, we estimate R= 0.95~R$_\odot$ for $\xi$~UMa~B.  IUE
spectra from both components (Table~\ref{tabiue}) show that most lines
are stronger in the B component of the system by a factor
$\sim$2:1. Hence it is expected that the B component will be the main
source of flux in the EUV wavelengths, but the A component is likely
to have a non-negligible contribution of EUV light.

The light curve (Fig.~\ref{plc}) has been phased using the ephemeris
given by \citet{griffin98} for the A and B components, and adapted to
follow the criteria of $\phi_{orb}$=0 corresponding to the primary
star behind, resulting in T$_0$(HJD)=2,442,442.916, P$_{orb}$=3.980507
d.  The DS light curves of $\xi$~UMa show some flaring activity, as
well as semi-periodic fluctuations similar to those observed in other
systems.

The \ion{Fe}{16} line ratio in the MW spectrum (2.25$\pm$0.15)
provides an accurate value of the hydrogen column density of
$N_H=8.\pm3\times10^{17}$~cm$^{-2}$ (Fig.~\ref{ismcalc}), lower than
the value of $N_H=1.5\times10^{18}$~cm$^{-2}$ estimated by
\citet{sch95} from nearby stars.  Fig.~\ref{emdfigs} shows the EMD
calculated for this system. This is the most outstanding example of
two peaks in the emission measure, with very similar values of the
emission measure for each peak. This result fits the observed fluxes
better than an EMD with only one peak at intermediate temperatures.

\subsubsection{$\alpha$~Cen}
As in the case of $\xi$~UMa, EUVE can not separate the light coming
from $\alpha$~Cen A (HD 128620, G2V) and B (HD 128621, K1V).  Hence,
the observed light curve and spectra correspond to both stars. IUE
spectra show higher flux levels in the A component (see
Table~\ref{tabiue}) than in the B component, but {\em EINSTEIN}
satellite x-ray observations show that the K1 star is the predominant
source in the range 0.2--4~keV \citep{gol82}, with an approximate
ratio 2:1.

The DS light curve of $\alpha$~Cen does not show any short-term
change, and no flares are registered. Analysis of seasonal changes
shows small variations, counting an increase in the flux by $\sim$15\%
from 1995 to 1997, while the 1993 observations are compromised by the
dead spot.  The hydrogen column density adopted is $N_H=6\times
10^{17}$~cm$^{-2}$, calculated by \citet{lin96} from Lyman~$\alpha$
and Mg~II~h~\&~k lines.  As in the case of Procyon, disagreements
occur between the observed and predicted fluxes in the \ion{Fe}{12}
$\lambda$364.4.

The general shape of the EMD is similar to that of the Sun in the
absence of flares, up to log~T(K)$\sim$6.5 \citep*[see][]{rea01}.  The
electron density calculated from \ion{Fe}{10}, \ion{}{12}, \ion{}{13},
and \ion{}{14} lines in the range log~T(K)$\sim$6.0--6.5 gives a value
of log~$N_e(cm^{-3})\sim$9.5 (Mewe et~al. 1995; Drake, Laming, \&
Widing 1997).

\subsection{Active stars}

This group of stars includes those for which the EMD 
is clearly dominated by
the emitting material at log~T(K)$\sim$6.9 signaled by strong
lines of \ion{Fe}{18} and \ion{Fe}{19}, but with relatively small
contribution from material at higher temperatures (compared to the
most active stars in the third group).  Capella \citep{dup02} is 
included in this group, and was also present in the sample of 28 stars
studied in \citet{tesis}.

\subsection{$\beta$ Cet}
$\beta$~Cet (HD 4128, HR 188, K0 III) is a single star with an
apparent low rotational velocity \citep*[v$\cdot$sin~i= 3.5 $km\
s^{-1}$][]{melo01}, but with surprisingly high levels of activity in
X-rays. An estimate of the rotational period of 189.1~d can be deduced
from the radius and inclination in Table~\ref{tabparam} and
v$\cdot$sin~i \citep{gray89}.  Two sets of EUVE observations are
available for $\beta$~Cet. An observation of 6 days during 1994
September shows a light curve with no significant variation, and a
spectrum and EMD similar to that of active binary systems like Capella
\citep{paper1}.  A second set of observations taken during 2000 August
shows flaring events in the DS light curve, and a level of emission
much higher than in the 1994 campaign suggesting seasonal changes in
the level of coronal activity of this star \citep*{ayr01}. The
accumulated 808~ks of exposure time in the SW spectrum makes it the
longest stellar observation with EUVE, and allows the analysis of a
high-quality spectrum.\footnote{ Unfortunately detectors for the MW
and LW spectrometer were not turned on for this exposure.}
\citet{ayr01} applied models based on upper-limit EMDs to the observed
spectra of quiescence and flaring stages during this set of
observations\footnote{The observations in the EUVE archive contain one
set of corrupted data, and the first $\sim$4 days of observation
included in \citet{ayr01} are not available in the archive.}. These
authors report a total exposure time of 645~ks, well below the
exposure time calculated from archival data.  This affects the fluxes
reported by \citet{ayr01}, which are larger than fluxes reported here
by $\sim$15-35~\%.  No details are given by these authors on how well
their model predicts the line fluxes, and during flaring stages their
model spectrum seems not to accurately match some of the observed
lines, including the density sensitive lines.

The EMD derived from  the summed 2000 spectrum yields  one of the best
fits of the sample (see Fig.~\ref{emdfigs}), although the lack
of MW and LW spectra containing the \ion{Fe}{15},
\ion{Fe}{16} and \ion{Fe}{24} lines, weakens the constraint on the
EMD at log~T(K)$\sim$6.5 and log~T(K)$\sim$7.3.
The resulting EMD shows clear differences with respect to the
1994 observations in the temperature range available in this
observation, exhibiting a higher EMD mainly for temperatures higher than the
bump at log T(K) $\sim$ 6.8 \citep[see][]{paper1}. 

\subsubsection{AY Cet}
AY~Cet (HD 7672) is an RS CVn system with the EUVE flux dominated by a
G5III active star, and a faint white dwarf companion with negligible
flux contribution in this band. 
\citet{sch95} had difficulties applying a global fit to the noisy
spectrum of AY~Cet, and only the assumption of a low iron abundance 
allowed a result
without a false ``hot tail'' in the EMD. The global
fit yielded  an EMD
peaking at log~T(K)$\sim$7.0. Our EMD modeling
indicates a peak at  log~T(K)$\sim$6.9 (Fig.~\ref{emdfigs}) with no
``hot tail.'' The shape of our EMD does not depend on abundances,
since we use only Fe lines.

\subsubsection{AR Psc}
AR~Psc (HD 8357, G7 V + K1 IV) has a shorter photometric period
\citep*[P$_{phot}$=12.38 days;][]{cuti01} than the orbital period of
14.3023 days \citep{fek96}, something unusual among RS~CVn
stars. \citet{fek96} proposed that AR Psc has not yet arrived on the
main sequence to explain the lack of synchronization between the two
periods. The EUVE observations of AR~Psc with the DS (Fig.~\ref{plc})
reveal the persistence of an active region during a full rotational
period.  This active region is visible around JD$\sim$2,450,690.2 and
JD$\sim$2,450,702.5, approximately the duration of the optical
photometric period.  Enhanced emission is also observed after the
second appearance of this active region; this feature can not be
unambiguously considered as a flare since the low enhancement of light
($\sim$50\%) does not match the duration of the event ($\sim$1 day)
when compared to other flares of similar duration
\citep{ost99,paper1}.\footnote{\citet{ost99} report the presence of two
flares in the AR~Psc light curve, with the first flare starting at
phase $\phi_{orb}$=1.082.  This is an orbital phase when no data were
taken and does not correspond to the date reported by these authors.
In the second flare, at $\phi_{orb}$=1.378, 
the beginning of the second enhancement that we identify,
a rise time of 51.3~hr and a decay time of 42.1~hr are reported
as e-folding times, but these times do not match the duration of the
observed rise and decay of the flare.
}
  
There is much uncertainty in the hydrogen column density used for this
star. A value of $N_H=2\times10^{18}$~cm$^{-2}$ is assumed, estimated
from column densities measured in nearby stars \citep{fru94}.  AR Psc
shows a fairly small bump in the EMD (Fig. \ref{emdfigs}), but peaking
at a higher temperature than usual, at log~T(K)$\sim$7.1. A better
knowledge of the hydrogen column density is needed to specify the EMD
particularly in the lower T regions defined by the iron lines at long
wavelengths.

\subsubsection{CC Eri}
Similar to other systems observed in this sample, small-scale
variability is  present in the light curve of 
CC~Eri (HD 16157, K7 V + M3 V),
with non-periodic variations of 10--14~hr (Fig.~\ref{plc}).   
A short-duration
flare could be present at JD$\sim$2,449,977.5.
A value of $N_H=2.6\times10^{18}$~cm$^{-2}$ was estimated by 
\citet{pan95}
from the distance to the star and the average hydrogen column density.
\citet{ama00}, using this value, derived an
upper limit to the  EMD based on the   
temperature of maximum emissivity of IUE and EUVE spectral lines. Their
derived  EMD has a minimum around log~T(K)$\sim$5.0 and a peak at
log~T(K)$\sim$6.8.   

In the EMD we calculate with the whole emissivity function, the 
resulting   
distribution (see Fig.~\ref{emdfigs}) shows a minimum  at
higher temperatures (between 5.5 -- 5.8 dex), and 
an apparent  overabundance of
nitrogen is indicated by the \ion{N}{5} $\lambda$1240 line, similar to
the cases mentioned in \citet{paper1}.  The local enhancement in
the EMD occurs at log T(K)= 6.8.

\subsubsection{YY Gem}
YY~Gem (Castor C, HD 60179C) is a well known active system with two M
dwarf stars in the double-lined spectrum.  The DS light curves are
characterized by the presence of many short flaring events (at least
7), including an enhancement by a factor of at least $\sim$9 at the
very end of the observation (Fig.~\ref{plc}).  Lower flux levels are
displayed in the right panel to show the small-scale variability. This
system has an orbital inclination of 86.3$\arcdeg$ (see
Table~\ref{tabparam}), making it one of the best targets to search for
rotational modulation and eclipses, as have been found with ROSAT
\citep{schm98}. But the presence of frequent flaring, along with
the noise found in the data, makes it difficult to find such evidence
from the present observations.

The hydrogen column density towards nearby stars \citep{fru94}
suggests a column density of $N_H \sim 2.5\times 10^{18}$~cm$^{-2}$ to
YY Gem, but this value seems to be high compared to nearby stars in
our sample.  The use of the \ion{Fe}{16} line ratio
($\lambda$335/$\lambda$361) in the MW (1.8$\pm$0.3) yields an upper
limit to the column density of $N_H \la 6\times 10^{17}$~cm$^{-2}$
(Fig.~\ref{ismcalc}). In view of these divergent results, we use the
upper limit given by the \ion{Fe}{16} line ratio, since it represents
a direct measurement towards this star.

The EMD of the system (see Fig.~\ref{emdfigs}) can be used to predict
the \ion{Ar}{15} $\lambda$221.15 line using APEC and good agreement is
found, assuming [Ar/Fe]~0.8 (a noble gas enhancement) 
and  a solar oxygen abundance \citep{anders}.
Also, the complex blend at $\lambda$192~\AA\ includes the
\ion{Fe}{12} $\lambda$193.51 line, which may contribute $\sim$40\% of
the observed flux. These lines give some information regarding the
stellar EMD near log~T$_e(K)\sim$6.3, although a better determination
of the ISM absorption would be helpful.

\subsubsection{BF Lyn}

Variations by up to $\sim$20\% with respect to the average value are
observed in the light curve of BF~Lyn (HD 80715, K2 V + dK),
suggesting the presence of small-scale modulation in a semi-periodic
pattern.  As in the case of YY~Gem, the value of the hydrogen column
density towards BF~Lyn is very uncertain, although the presence of the
\ion{Fe}{16} lines makes possible an estimate from their ratio
(1.54$\pm$0.63) of $N_H \la 5\times 10^{17}$~cm$^{-2}$, as shown in
Fig.~\ref{ismcalc}. Since the S/N of these lines is quite low, an
intermediate value to that reported in nearby stars (similar to those
in the case of YY~Gem) has been adopted.  The assumed value in the EMD
calculations is $N_H=1.5\times 10^{18}$~cm$^{-2}$, more consistent
with the EMD shape estimated when lines at long wavelengths are
excluded from the fit.

\subsubsection{LQ Hya}
LQ Hya (HD 82558) is one of the single stars included in the
sample. The youth of this star appears to be the main cause of the
high levels of activity observed \citep[see][ and references
therein]{mon99}. The EUVE light curve (Fig.~\ref{plc}) shows
variations by a factor of 2 during the ``quiescent'' state of the
corona, and also two impulsive flares (intense flares of short
duration) are present.  The variations of the ``quiescent'' corona do
not follow a clear periodic pattern, and are not related to the
photometric period of 1.63~days \citep{cuti01}, but optical modulation
is present on a scale of $\sim$1.2 days.

\citet{woo00} find an abnormally high value of the hydrogen column
density of $N_H= 1.1^{+0.5}_{-3.1}\times 10^{19}$~cm$^{-2}$, from an
analysis of Lyman~$\alpha$ and Mg~II~h~\&~k lines.  Since the Mg lines
could be affected by stellar activity and and circumstellar gas, and
data from nearby stars seems to disagree strongly with these values,
we used the conservative value given by the lower limit, $N_H=8\times
10^{18}$~cm$^{-2}$.  Given the lack of lines at longer wavelengths in
the spectrum of LQ Hya (that could be affected even more by
uncertainties in the adopted value of $N_H$), changes in the shape of
the EMD due to such uncertainties will be minimal, leading mainly to a
vertical displacement of the EMD.  Good agreement is obtained between
the predicted and observed fluxes (see Fig.~\ref{emdfigs}).

\subsubsection{DH Leo}
The observed 1995 light curve of DH~Leo (HD 86590) shows much
variability (changes by up to $\sim$37\% from the average value,
Fig.~\ref{plc}), and some short flare-like events
occur as well. \citet{ste96} proposed that rotational modulation could be
present in these observations.  The \ion{Fe}{16} line ratio does not
provide an accurate value of the column density in DH~Leo since the
361\AA\ line has poor statistics. But the low flux observed in this
line points to a rather high column density. Values of nearby stars in
\citet{fru94} suggest $N_H \sim 1\times 10^{18}$~cm$^{-2}$. On the
other hand, \citet*{diam95} deduced a wide range of values of
$N_H=4.^{+28}_{-N/A}\times 10^{18}$~cm$^{-2}$ obtained from a fitting
to ROSAT spectra. An intermediate value with those of nearby stars has
been adopted, $N_H\sim 2\times 10^{18}$~cm$^{-2}$.

\subsubsection{BH CVn}
BH CVn (HD 118216, HR 5110) shows a remarkable pattern of variability
in the EUVE light curve (Fig.~\ref{plc}), with semi-periodic
variations of $\sim$20--30~hr.

There is much uncertainty in the determination of the hydrogen column density
in the direction of BH~CVn. Spectral fits to ROSAT data yield very
high column densities 
(over $N_H \ga 1\times 10^{19}$~cm$^{-2}$, Diamond et~al. 1995;
{Graffagnino}, {Wonnacott}, \& {Schaeidt} 1995),
although this is not a reliable
method to determine $N_H$. The \ion{Fe}{16} line ratio of
1.45$\pm$1.17 does not point to an accurate value either, due to the
low S/N of the lines, but the upper limit of this ratio corresponds to
$N_H \sim 3\times 10^{18}$~cm$^{-2}$ (Fig.~\ref{ismcalc}). On the
other hand, the values derived from nearby stars \citep{fru94} yield a
lower estimate of $N_H \sim 7\times 10^{17}$~cm$^{-2}$. As a
compromise among these values we have chosen the upper limit given by
the \ion{Fe}{16} line ratio, $N_H= 3\times 10^{18}$~cm$^{-2}$, 
also consistent with \citet{mit97}.  The
calculated EMD (Fig.~\ref{emdfigs}) shows a bump similar to other
stars, with a decreasing EMD at higher temperatures.

\subsubsection{V824 Ara}
About 2.5 days of DS light curve observations were obtained for
V824~Ara (HD 155555) in 1996. The low flux (note the 3000~s binning)
gives quite large error bars (see Fig.~\ref{plc}). Nevertheless, it is
possible to identify some modulation coincident with the orbital and
photometric periods of the system ($\sim$1.68 days).  Local maxima
appear at phases 1.0, 1.5, and 2.0, consistent with orbital
modulation, although flare-like variability can not be excluded.

The small number of lines present in the spectrum of this young active
binary system makes the estimate of the EMD less accurate.
\citet{aira98} made an analysis based on the peak of the emissivity
function in order to calculate minimum values of the emission
measure. The resulting EMD (Fig.~\ref{emdfigs}) shows a poorly
constrained bump near log~T(K)$\sim$6.9. The electron density
(Table~\ref{tabne}) derived from the ratio \ion{Fe}{21}
$\lambda$102.22/$\lambda$128.73 is not very reliable, since the
$\lambda$102.22 line flux was measured including a blend with the
\ion{Fe}{19} $\lambda$101.55 line. The flux of the \ion{Fe}{19} line
was estimated from the EMD to account for its contribution to the
blend.

\subsubsection{ER Vul}
The DS light curve of the partially eclipsing system ER~Vul (HD
200391) shows some variability, as well as flare-like activity at a
low level. \citet{ost99} suggested the variations resulted from
small-scale flares, and found no eclipses or periodicities in this
BY~Dra-type system.   The EUVE observations span
$\sim$10 epochs of the binary enhancing detection of
periodicity.  A power spectrum of  the DS light curve 
(with flaring portions omitted) shows a maximum corresponding to the
optical period.  When the DS light curve is phased to this period,
there is modulation at most by a factor of 5 with a suggestion of 
absorption near phase 0.5  when the G0V component is partially
occulted by the G5V star.  Thus the corona may be
more compact in the hotter star of ER~Vul.

There is much uncertainty in the value of the hydrogen column density
towards ER Vul, since the values towards nearby stars reported in
\citet{fru94} show discrepancies.  The value adopted by \citet{ruc98}
of $N_H=3\times 10^{18}$~cm$^{-2}$ has also been used in this
work. This uncertainty could affect the determination of the slope at
temperatures below the bump in the EMD because that region is defined
by lines at longer wavelengths where the impact of interstellar
absorption is largest.

\subsubsection{BY Dra}
The DS light curve of BY Dra (HD 234677) covered an orbital
period of this system (Fig.~\ref{plc}). 
Although some variation occurs that could 
correspond to the rotational period of the system of $\sim$3.83 days (see
Table~\ref{tabparam}), several flares  prevent a clear 
identification of rotational modulation.

Values from nearby stars suggest a quite high hydrogen column
density ($N_H=5\times 10^{18}$~cm$^{-2}$), 
consistent with the lack of flux detected in the MW spectrum,
probably due to ISM absorption. We have used this 
value in the absence of other evidence.  Unless large differences 
in the hydrogen column are present with respect to this 
assumed value, there will be only a minimal influence on the 
shape of the EMD (Fig.~\ref{emdfigs}).

The electron density calculated for BY Dra makes use of lines
with S/N lower than 3, and hence these values are not very reliable.
The EMD is fit to emissivities at log~N$_e$(cm$^{-3}$)$\sim$12, which
gives a good fit to the density-sensitive resonance line \ion{Fe}{21}
$\lambda$128.7.

\subsection{Very active stars}
Some stars in the sample contain substantial amounts of material at
temperatures beyond log~T(K)$\sim$7.0, sometimes with 
emission measures even larger than the
values at log~T(K)$\sim$6.9.  Most of those stars 
included in \citet{paper1}, UX~Ari, V711~Tau, $\sigma$~Gem, II~Peg, and
AB~Dor belong in this category of high activity. The 
presence of material at these temperatures is 
determined by the \ion{Fe}{23} and \ion{Fe}{24} lines.
\ion{Fe}{24} occurs in the MW spectrum which frequently suffers
from insufficient exposure time in current 
data sets.  Future observations, for instance with
Chandra or XMM-Newton,  could reveal high temperature
emission that was not detected in short exposures with EUVE. 

\subsubsection{VY Ari}
The DS measurements of  VY~Ari (HD 17433)  are compromised 
by the dead spot of the DS detector during the observation 
(Fig.~\ref{plc}). 
However, the SW light curve supports the modulation observed in
  the DS light curve,
  with a decrease of flux of $\sim$60\% during the observation.

We have used the same value for the  hydrogen column density as in the
case of the nearby star UX~Ari
\citep[$N_H=1.5\times10^{18}$~cm$^{-2}$,][]{paper1}.
In constructing the EMD, we find that
the predicted fluxes
in two of the \ion{Fe}{20} lines do not well match the measured values.
While the observed flux of $\lambda$110.63 is too strong,
the $\lambda$121.83 transition  
is too weak. These lines could be affected by other blends
not included in the analysis.
In any case, this makes  the electron 
density (cm$^{-3}$) of 13.8 dex deduced from the 
ratio \ion{Fe}{20} $\lambda$110.63/($\lambda$118.66+$\lambda$121.83) 
less certain, and  points towards a lower
value, probably closer to that deduced from
\ion{Fe}{21} ratios of $\sim$12.5 dex (see Table~\ref{tabne}).
The resulting EMD (Fig.~\ref{emdfigs}) is not very sensitive to the
use of a different electron density in this range (the adopted value
was log~N$_e$[cm$^{-3}$]$\sim$13.0).

\subsubsection{$\sigma ^2$~CrB}

The DS light curve of $\sigma ^2$ CrB (HD 146361) shows intense flares
(Fig.~\ref{plc}), and also small-scale variability on the order of
several hours ($\sim$6--10~hr). The flare timing of this observation
was analyzed in detail by \citet{ost99} and \citet{ost00}.

The good S/N achieved in the \ion{Fe}{16} lines in the MW spectrum
allows an accurate determination of the hydrogen column density,
resulting in a value of $N_H=2.5^{+1.5}_{-0.9}\times
10^{18}$~cm$^{-2}$ (Fig.~\ref{ismcalc}), consistent also with
  \citet{mit97}.  Fig.~\ref{emdfigs} shows an 
EMD that reflects a very hot corona, with an increasing value at
temperatures higher than the ``bump.''  Also the general level of the
EMD is comparable to that of RS~CVn systems composed of subgiants,
whereas the components of $\sigma^2$ CrB are dwarf stars,
demonstrating their unusually high activity levels.

\subsubsection{V478~Lyr}
The lack of good statistics in the DS light curve (Fig.~\ref{plc}) of
V478~Lyr (HD 178450) masks possible low level variability.  Variations
of $\sim$25\% with respect to the average flux are found in the data,
with no clear evidence of DS eclipses. The system has an inclination
of 83$\arcdeg$ and undergoes partial eclipses \citep{cabs}.

An intermediate value of the hydrogen column density, $N_H=4\times
10^{18}$~cm$^{-2}$ has been used from estimates for stars nearby in
the sky \citep{fru94}.  The  EMD is not affected
significantly by the uncertainties in $N_H$ (see
Table~\ref{tabparam}), since all the emission line fluxes are taken
from the SW spectrum, where interstellar optical depths are small.
Although the number of lines measurable with reliable statistics in
this system is small, it was possible to construct an EMD similar to
other stars in the very active group.

\subsubsection{AR Lac}\label{arlac}
AR~Lac (HD 210334) is an eclipsing binary with two active stars
(G2IV/K0IV) that shows chromospheric emission originating from the
K0IV star \citep{mon97}. The measurement of the primary eclipse depth
in the 1993 and 1997 observations (Fig.~\ref{plc}) reveals a
contribution by the K0IV star of at least a $\sim$37\% (measured at
the center of the primary eclipse) of the total EUV light.  The
partial contamination of the light curve by the dead spot in 1993, and
the presence of flares around phases x.5 in 1993 and 1997 prevent a
reliable measurement of the secondary eclipse.  The light curve in the
2000 observations, reported by \citet{pea01}, is dominated by a large
flare.  The EUV flux increases by at least a factor of $\sim$17 from
the quiescent level. A total net energy release of 2.0$\times
10^{35}$erg is found in the range 80--170~\AA, after the subtraction
of the ``quiescent'' contribution of the 1993 and 1997 summed
observations, following the method explained in \citet{paper1}.  The
flux obtained before the subtraction of the quiescent contribution was
4.4$\times 10^{35}$erg. A complementary analysis of the observations
in 2000, with partially simultaneous Chandra data can be found in
\citet{huen01}.

During the 2000 flare, AR~Lac shows an increase in flux in the hottest
lines of the EUVE spectrum, and some increase also in the continuum
level.  There may be an increase of the average density measured
during the flaring observation as indicated by \ion{Fe}{22} but not
confirmed by other ratios (see Table~\ref{tabne}).

Although there is some uncertainty in the measurement of the
\ion{Fe}{16} lines (especially the $\lambda$360.8 line), we use the
hydrogen column density derived from their flux ratio (2.5$\pm$0.8)
found in the MW spectrum corresponding to the 1993 and 1997
observations co-added (Fig.~\ref{ismcalc}).  The resulting value
$N_H(cm^{-2})=1.8\times 10^{18}$ is consistent with the column density
of $N_H(cm^{-2})=2\times 10^{18}$ assumed by \citet{gri98}, calculated
from nearby stars.

\citet{kaa96} and \citet{gri98} used the 1993 data to obtain EMDs with
a clear peak around log~T$_e(K)\sim$6.9, although they differ in the
distribution at other temperatures. A preliminary analysis of the 1997
data alone was made by \citet{bri99}. The addition of the 1993 and
1997 data allows the analysis to be extended to almost the whole range
of log~T$_e(K)\sim$4.0--7.4.  The 2000 data are used to compare a
flaring sequence to the ``quiescent'' observations of 1993 and
1997\footnote{Although these observations contain some flaring
activity, they are dominated by the quiescent state.}.

Fig.~\ref{emdfigs} displays the EMD calculated from the 1993 and 1997
summed data.  The height of the bump is less prominent during the
flare because the material at higher and lower temperatures increases.
However the value of the EMD at the bump remains constant
demonstrating its stability in the corona (Fig.~\ref{stagesarlac}).
This reinforces the hypothesis of large flares as phenomena unrelated
to the bump at log~T$_e(K)\sim$6.9, suggested previously by
\citet{paper1}.

\subsubsection{FK Aqr}
The nearby active binary system FK Aqr (HD 214479) is composed of two M
dwarf stars orbiting with a $\sim$4 day period. 
The DS light curves in the 1997 campaign show frequent 
flaring activity  (Fig.~\ref{plc}), with impulsive flares of 
short-duration ($\sim$15~hr) that are quite strong 
(increases by up to a factor of 5). The presence of these 
flares blurs any possible
modulation related to rotation. In contrast, the 1994 observations
show a quiet corona, and only a small
flare is present. Some enhancement of flux arises during the
second half of the orbital period in 1994, probably due to the presence of an
active region in the line of sight, as \citet{ste96} proposed.
Values of inclination and photospheric radii of this system can be
inferred in a first approximation by assuming that the
pair of M2V/M3V dwarfs follows the relations derived by \citet{gray92}. 
This would imply stellar radii of 0.55/0.52~R$_{\odot}$, 
and with the typical
mass of 0.42 and $m\, sin^3 i=0.27$ for the primary, an inclination of
i$\sim$60$\arcdeg$ can be estimated. 

There are no direct measurements of
the interstellar column density towards this system, 
but stars nearby in the sky \citep{fru94}
suggest
a value of $N_H=7-10\times 10^{17}$~cm$^{-2}$. The \ion{Fe}{16} line
ratio in the MW spectrum (1.96$\pm$0.20) provides an upper limit to
the hydrogen column density of $N_H=7\times 10^{17}$~cm$^{-2}$
(Fig.~\ref{ismcalc}). This 
value was assumed here.
Fig.~\ref{emdfigs} shows a remarkably well-fitted EMD, with a first
bump in temperature similar to the solar EMD \citep[see][]{rea01},
and the typical second bump found in most of the stars in
this sample around log~T$_e(K)\sim$6.9.

\section{Discussion}\label{sec:discussion}

One goal of this research is to identify the stellar 
parameters that influence the coronal structures.
Systematics of the light curves, EMD, and densities are
discussed below. We then compare the  underlying patterns of coronal
structure to the properties of the stars.

\subsection{Light Curves}
The intrinsic variability found in the light curves of these systems
makes it difficult to confirm seasonal (year-to-year) changes in the EUV
emission. Among the quiet stars, only $\alpha$~Cen shows some small
decrease in the light curve flux (by $\sim$15\%) between 1995 and
1997, but the observations are too short to attribute these changes to
seasonal variations. The fluctuations found in the quiescent emission
of II~Peg are of order $\sim$50\% \citep{paper1}, although in this
case the effects of intrinsic variations could be more important. The
clearest case for seasonal fluctuations is found in the single giant
star $\beta$~Cet, which in 1994 is very quiet for the 6.4 day
observation. The DS flux is about a factor of 2 lower than in 2000
\citep{paper1}, and \citet{ayr01} identify at least 5 flaring events
during the $\sim$34 days of monitoring in 2000 (see Fig. 1).  Some
line fluxes are enhanced by more than 50\% (a factor of 5 for
\ion{Fe}{23}/{\sc \small XX} $\lambda$132.85!). The intrinsic
variability found in stars like $\sigma$~Gem, LQ Hya or FK Aqr, even
in the absence of large flares, makes it difficult to confirm seasonal
changes.

Signatures of rotation are clearly found in the eclipses of AR~Lac,
and might be present in the light curves of AR~Psc, ER~Vul,
  VY~Ari and BY~Dra.
Marginal evidence for modulation was found previously in II Peg, UX
Ari, $\sigma$ Gem, and AB Dor \citep{paper1}.  Additional short-term
variations are found in many stars in this sample, indicating
non-periodic changes on time scales between 0.3 and 1.5 days.
Instrumental effects can be rejected in the case of V711~Tau, and
these variations are likely to be real for many stars in the sample,
including V711~Tau, LQ~Hya, $\sigma^2$~CrB and BF Lyn (see
Fig.~\ref{plc}).  This variability can be attributed to flare-like
phenomena on a small scale, perhaps reflecting the existence of many
solar-like flares (shorter duration and intensity than the large
flares which are well observed in these stars).

\subsection{Emission Measure Distribution}

A continuous distribution of material occurs in these coronas,
spanning 3 decades of temperature or more.  It is now clear that the
earlier simplifications of 1-T or 2-T coronal models resulted from
insufficient spectral resolution and/or incomplete global models
preventing identification and analysis of the wide range of excitation
and ion stages naturally present in the coronae of cool stars.
Results assembled in Fig. 4 show the distribution of material to be
generally similar among active cool stars, whether single or binary.
A decrease in the EMD between 10$^4$K and 10$^{\sim 5.5}$K and then an
increase, with one or more local enhancements between 10$^6$ and
10$^7$K represent the structure of the stellar chromosphere,
transition region, and corona.

However, a detailed comparison shows that clear differences appear at
temperatures above log~T(K)$\sim$5.2.  Only $\alpha$~Cen, with a
solar-like EMD, has a distribution that begins increasing above
temperatures as low as log~T(K)$\sim$5.2.  The remaining stars in the
sample have the minima in the EMD at higher temperatures
[log~T(K)$\sim$5.8]. This difference between the minima of the Sun and
Capella was noted from the first EMD derived from EUVE \citep{dup93}.
  We note that the Sun shows minima at different
temperatures for different types of structures. For example, solar
coronal holes frequently have a higher temperature EMD minimum than do
active regions.

At temperatures above 10$^6$ K, additional structures are found,
characterized by local enhancements of the EMD over narrow temperature
ranges.  Such a feature was also discovered in Capella and labeled the
``bump'' \citep{dup93,bri00}, and have been
identified with high latitude coronal features \citep{bri98}.  Several
stars show a local enhancement at log T(K)=6.2 
reminiscent of the temperature of the solar corona in addition to a
bump at log T(K) =6.8.  The presence of this second high temperature
bump is unambiguously found in 25 out of 28 stars in the sample (all
except $\epsilon$~Eri, $\alpha$~Cen, and Procyon).  Clearly, these
represent a fundamental coronal structural feature.  Finally, the
stars with an increasing EMD beyond log~T(K)$\sim$6.9 represent the
very active classification.  The progressive addition of hotter
material marks the increase in activity level.

Given the different physical sizes of the stars in the sample, it is
useful to evaluate the emission measure weighted by the emitting
surface of the stars (4$\pi$[R$_1^2$+R$_2^2$]), so that the
emission measure per unit area defines an effective column density of coronal
material. 
Fig.~\ref{sevstars} shows 6 cases representative of
different degrees of activity, weighted by the size of the emitting
stars -- the ``column'' EMD.   Procyon  shows the same amounts of
material as $\alpha$~Cen at temperatures above log~T(K)$\sim$6. 
Except for $\alpha$~Cen, 
the chromospheric structures are similar, while the greatest
divergence occurs at the highest temperatures.

\subsection{Electron Density}
Electron densities derived from line ratios at log~T(K)$\sim$7.0
indicate values of\\ log~$N_e(cm^{-3})\ga$12. There is considerable
dispersion in the data for values of log~$N_e(cm^{-3})\ga$13, as noted
previously for Capella \citep{bri96}, as well as for densities at
lower temperatures in solar active regions \citep{bri95}. Numerous
observational issues compromise these results. The presence of blends
not well evaluated in the models and uncertainties in the placement of
the continuum used as a base for the line flux measurements clearly
influence some of these measurements.  In particular, the \ion{Fe}{20}
line ratios show a systematically higher density than other iron line
ratios at similar temperatures, as well as a poorer fit of the
emission measure; therefore more caution must be taken with the
results from this ion. While the atomic models for the diagnostic line
ratios have been benchmarked under controlled conditions
\citep{four01}, higher resolution spectra with good signal to noise,
are needed.

Nevertheless, density diagnostics from several stars in this sample
have good statistics and consistency among several diagnostics,
e.g. VY~Ari and CC~Eri discussed here and V711~Tau and $\sigma$~Gem
\citep{paper1}.  Given these results, a ``conservative'' value of
log~$N_e(cm^{-3})\sim$12 seems plausible.  We note that lower values
of density at log~T(K)$\sim$6.2 found by several authors in different
stars \citep[see \S~\ref{sec:lowactive} and, e.g.][]{can00} are not
necessarily inconsistent. \citet{bri96} used similar results for
Capella from Fe M-shell and L-shell diagnostics, and suggested the
presence of two different types of structures.

The densities calculated at log~T(K)$\sim$7.0, in combination with the
emission measure values at that temperature, can be used as a first
approximation to estimate the scale size of the emitting 
structures \citep{sanz01}, with the caveat that there is no
information on their geometry and filling factors. The
calculations show in all cases that such structures 
are small ($\la$0.02~R$_*$), both for dwarfs and giant stars.
Small structures have been found also by several authors, e.g.,
  \citet{dup93},  \citet*{bow00}, \citet{phi01} and references
  therein.

Current loop models cannot accommodate the presence of the high electron
densities found here at coronal temperatures. Observations of the 
Sun only detect such high 
values during solar flares \citep{phil96}.  Light curves from EUVE 
do not have enough signal in short integration periods
to detect the presence of fast solar-like flares in stars, so it is 
not possible to differentiate
such flares from the emission in quiescence. The presence  of frequent solar-like 
flares in these stars can not be ruled out or confirmed with
the present data.  Such flaring could account for both the
shape of the EMD and the high electron densities. 
A model of (continuous) nano-flare heating in the Sun can predict the shape
of the EMD with a bump at log T(K)=6.8 \citep{kli01},
but with densities two orders of magnitude lower than observed here.

\subsection{Comparison with Stellar Properties}

For a quantitative comparison of these EMDs, we extract
parameters of the distribution for comparison to physical
properties of the systems.  To define the temperature of
the peak of the EMD, we consider
the three largest values of the emission measure in
the range log~T(K)=5.8--7.3.
The temperature defined for the peak and its emission measure
were compared to the orbital
periods\footnote{The optical photometric period was employed in the absence
of orbital period.} of 30 stars, as shown in
Fig.~\ref{periods}. This sample includes also the data from
Capella \citep{dup93}, 44~Boo \citep{bri98}, $\lambda$~And
\citep{sanz01}, the 6 stars (V711 Tau, II Peg, $\sigma$ Gem,
UX Ari, AB Dor, and $\beta$ Cet) in \citet{paper1}, and the Sun
during  solar maximum \citep*{orl00}. 

The temperature of the peak of the EMD (Fig.~\ref{periods}a) remains
relatively constant at log T(K)= 6.9 for the binary stars
and 3 single stars (AB Dor, LQ Hya, and $\beta$ Ceti).
AB Dor and LQ Hya are young rapidly rotating effectively
single stars. Beta Ceti appears as an anomaly with strong
emission  - an apparently 
single slowly rotating star - consistent
with its classification as  K0 III giant. It may be a 
clump star that experiences a regeneration of its magnetic
dynamo, or it may be oriented pole-on to our line of sight, 
and the rapid rotation is not observable.

The mean electron densities at log T(K)=6.9 for the sample
are shown in Fig.~\ref{periods}b.  Density values range between
10$^{12}$--10$^{13.5}$ cm$^{-3}$ with no systematic dependences
on binarity or orbital period for the systems with P$_{orb}<$20~d.
There is no evidence for the highest densities in longer period
systems, but only 3 objects are in that group. 

Values of the EMD at the peak temperatures are shown in
Fig.~\ref{periods}c and \ref{periods}d.  For binaries, the value of
the EMD increases with increasing 
orbital period, as dwarf stars tend to have shorter periods
in our binary sample, whereas the larger RS~CVn subgiants and
giants have longer periods. 
Scaling the emission measure at its peak value
by the areas of the stars reveals the decrease in ``column'' 
EMD with increasing orbital period.  There may be a saturation
of the ``column'' EMD at periods less than 2.3 days where a constant
value appears consistent with the data.

Line fits to the data are 
superimposed, indicating the best fit for all the objects,
\begin{displaymath}
log (EM_{peak}/area)=50.4-0.85\ log P_{orb}
\end{displaymath}

\noindent
and only those with period longer  than 2 days,  

\begin{displaymath}
log (EM_{peak}/area)=50.5-0.89\ log P_{orb}
\end{displaymath}

\noindent
where P$_{orb}$ is given in days, EM in units of cm$^{-3}$, and area
is defined by 4$\pi$[R$_1^2$+R$_2^2$] , with radius 
in solar units.  In all
cases a solar photospheric abundance was assumed. Since the absolute
value of the emission measure peak depends on the iron to hydrogen
abundance, this assumption requires testing. Furthermore, 
if there are substantial iron enhancements or depletions as a function
of coronal temperature, the EMDs will need to be reconstructed
accordingly.

We also plot in Fig.~\ref{lumperiod} the orbital periods of the
systems against the EUV (80--170~\AA) luminosity weighted
by the bolometric luminosity (L$_{EUV}$/L$_{bol}$). The EUV luminosity
was calculated 
from the integrated SW spectrum of the stars, corrected for the
instrumental effective area and interstellar absorption. The bolometric
luminosity results from the application of the bolometric corrections
available in \citet{flo96}. 
This plot may be affected by
uncertainties in the calculation of the bolometric luminosity. The
EUV flux may arise from one star in a binary, yet the magnitude ($V$)
and color ($B-V$) of the system are used to calculate the 
bolometric luminosity.  The results in Fig.~\ref{lumperiod}
show much dispersion.  However, the general behavior of
increasing $L_{euv}/L_{bol}$ with shorter period  and a 
possible saturation at periods $\sim$ 1 day or less
are consistent with studies of L$_X$/L$_{bol}$ of
active cool stars 
(cf. Walter \& Bowyer 1981; Pallavicini et al. 1981; {Fleming},
{Gioia}, \& {Maccacaro} 1989).
An increase of flux in X-rays is generally found for faster 
rotators, with the relation
becoming ``flat'' at some point at rotational periods between 1 and
10 days, marking a so-called ``saturation limit''.  In the sample
considered here, a simple linear relation can fit  the
data, although the stars with fastest rotation ($\la$ 2.3 d) 
follow a flatter distribution with  period. 

Our results  suggest the
presence of three kinds of structures,  at
temperatures of\\ log~T(K)$\sim$6.3, $\sim$6.8, and  log~T(K)$\ga$7.2,
that dominate the emitting coronae of cool stars. Loop 
models predict a shape
roughly similar to that deduced here from the combined UV and EUV
analysis \citep[cf.][]{paper1}. These models generally balance
radiative losses by a heating function, and conduction redistributes
energy along the loop.  An emission measure would increase
until the peak temperature of the loop, beyond which the amount
of material would drop drastically. The addition of loops at higher
temperatures can compensate for this drastic fall in the loop
emission measure in order to reproduce the observed stellar EMD.  

The classic view of static loops with fixed
cross-section gives an EMD that increases linearly 
in the high temperature region, with a predicted 
slope of 1.5. But more complex loops, with
expanding cross-section 
({Schrijver}, {Lemen}, \& {Mewe} 1989; {Griffiths}, N.~W. 1999; 
Hussain et~al. 2002)
can account for
larger slopes near the peak temperature of the loop, although
not with high electron densities. 
Low activity stars, like the Sun, Procyon, or $\alpha$~Cen,
would be dominated by solar-like loops, peaking at log~T(K)$\sim$6.3,
and with electron densities in the range
log~$N_e(cm^{-3})\sim$9--10.5 (Drake, Laming, \& Widing 1995; Mewe
et~al. 1995; Drake et~al. 1997).
Solar-like flares can
produce a bump in the EMD at temperatures around log~T(K)$\sim$7.1
\citep{orl00,rea01}. Stars such as Capella \citep{dup93} and FK~Aqr 
are dominated by structures 
with maximum temperature around log~T(K)$\sim$6.9 and
log~$N_e(cm^{-3})\ga$12.  Finally, only the most active stars show the
possible presence (not well constrained with EUVE data) of hotter loops
that could explain the observed emission of the hottest lines. These
hot loops may be directly related to the existence of large flares in
stars like 
UX~Ari, $\sigma$~Gem, V711~Tau and II~Peg \citep{paper1}. For the case
of AR~Lac presented here, the value of the  emission measure increases
during the 
flares at all temperatures, with only a slight
increase in the EMD slope at the hottest temperatures (see Table~\ref{slopes}).

It is significant that the high temperature enhancement
(the ``bump'') appears ubiquitous in the coronae of cool
stars.  Moreover the temperature of this enhancement is
almost the same in a wide variety of cool stars.  
It is not clear why this happens.  \citet{geh93} 
noted a very small ($<$3\%) 
inflection in a theoretical
radiative cooling curve near 6$\times$10$^6$K and suggested
it might account for stable coronal structures.  However these
considerations apply when radiation dominates over conductive losses and when the 
abundances are photospheric.  Both constraints may
not apply  in stellar coronae.  Additionally, changes in the
ionization equilibrium and atomic physics will impact such
small details of the cooling curves.  It is fair to conclude
that current theoretical models can not reproduce the observed
emission measure distributions with high densities.

\section{Conclusions}\label{sec:conclusion}

\begin{enumerate}

\item Emission measure distributions (EMD) were derived from EUVE
spectra of 22 active binary systems and 6 single stars. The
overwhelming majority (25) of the
stars in the sample  show an outstanding ``bump'' -- a local
enhancement of the EMD -- over a restricted temperature range. This
bump occurs near log~T$_e$~(K)$\sim$6.8--7.0. Its presence does not
depend on the luminosity class of the star or the activity levels
present,
and confirms a fundamentally new coronal structure.

\item The emission measure per unit area (``column'' EMD) increases 
towards shorter orbital periods, with a possible ``saturation'' effect
at periods less than ~2.3 days.

\item \ion{Fe}{19}--{\small\sc XXII} line flux
  ratios, formed at log~T$_e(K)\sim$7 and measured in the summed
  spectra for each star indicate high electron 
  densities (log~$N_e[cm^{-3}]\ga$12).  In conjunction with
  lower densities found previously at lower temperatures, these
  results provide additional evidence for different structures in 
  stellar coronae.

\item A  second local enhancement of the EMD 
peaking at log~T$_e(K)\sim$6.3 could
reflect the presence of solar-like loops in the corona of some stars
in the sample, and dominates the EMD in  $\alpha$~Cen AB,
$\epsilon$~Eri, and Procyon.

\item \ion{Fe}{22}--{\small\sc XXIV} line fluxes indicate the presence
of much hotter material at temperatures log~T$_e(K)\ga$7 in some stars
(VY Ari, $\sigma^2$ CrB, V478 Lyr, and AR Lac).

\item The derived EMDs suggest these stellar coronae  are 
composed of solar-like magnetic loops (peaking at
log~T$_e[K]\sim$6.3). The loops at log~T$_e(K)\sim $6.9 are not yet
understood from loop models. Loops peaking at
log~T$_e(K)\ga$7.2 may be related to large flares.

\item Fluctuations in the EUV light curve of many stars in
the sample are observed in a non-periodic 8--36~hr pattern, indicating
the existence of frequent low-level variability.

\end{enumerate}

\acknowledgments

This research is supported in part by NASA grants NAG5-7224,
NAG5-11093, 
and NAG5-3550 and CXC Contract NAS8-39073 to the 
Smithsonian Astrophysical Observatory.   JSF is
grateful for support in part by the Predoctoral Program of the 
Smithsonian Astrophysical Observatory, and the Real Colegio 
Complutense at Harvard University. 
This research has made use of the SIMBAD database, operated 
at CDS, Strasbourg, France, and of data obtained through 
the High Energy Astrophysics Science Archive Research Center Online
Service, provided by the NASA/Goddard Space Flight Center.  This
research has also made use of NASA's Astrophysics Data System Abstract
Service. The authors want to acknowledge D.~J. Christian
(CEA/Berkeley) for his aid with the management of the EUVE archival
data.

\clearpage
%

\setcounter{figure}{0}
\begin{figure}
\plotone{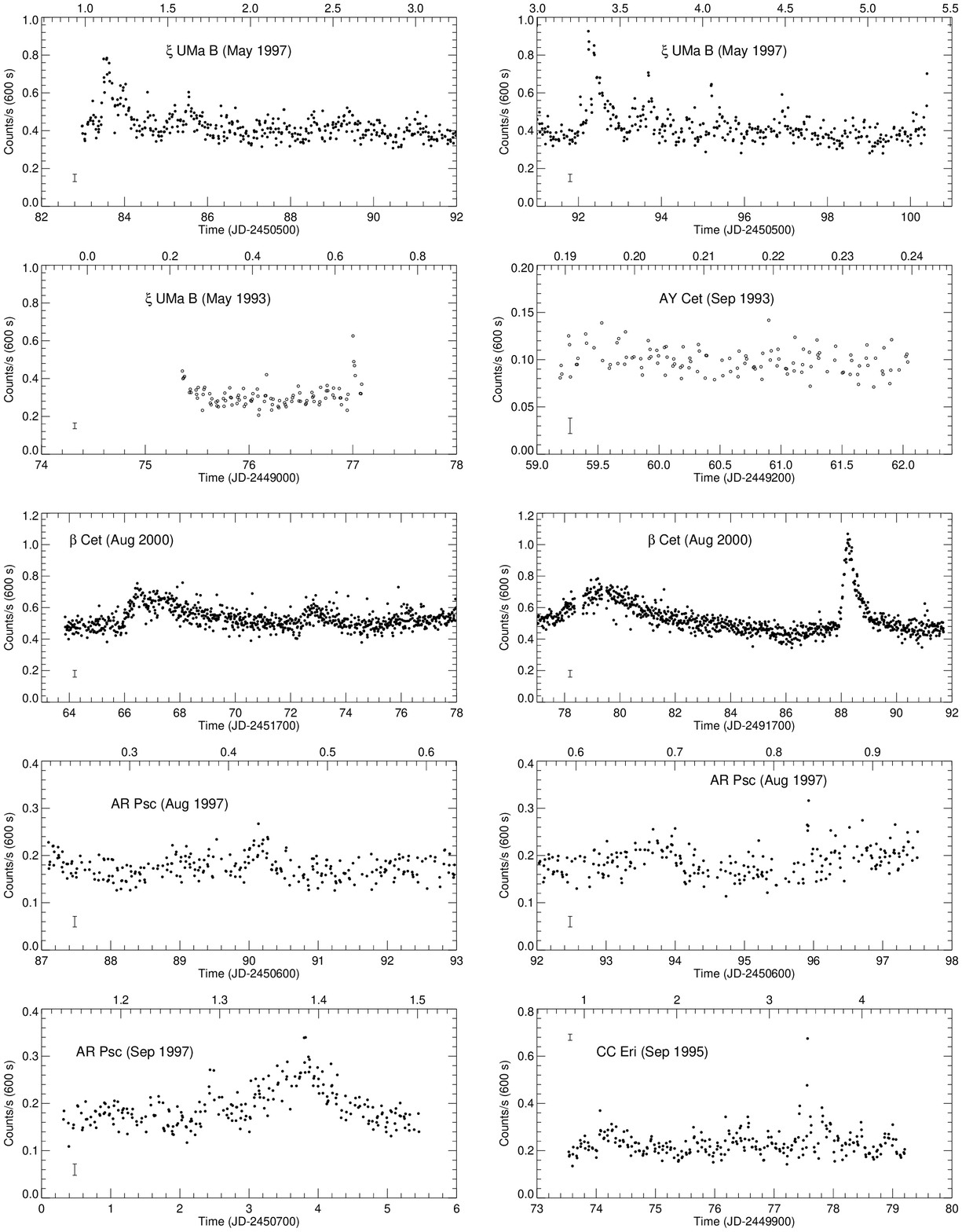}
  \caption{(b) continued.} 
\end{figure}

\setcounter{figure}{0}
\begin{figure}
\plotone{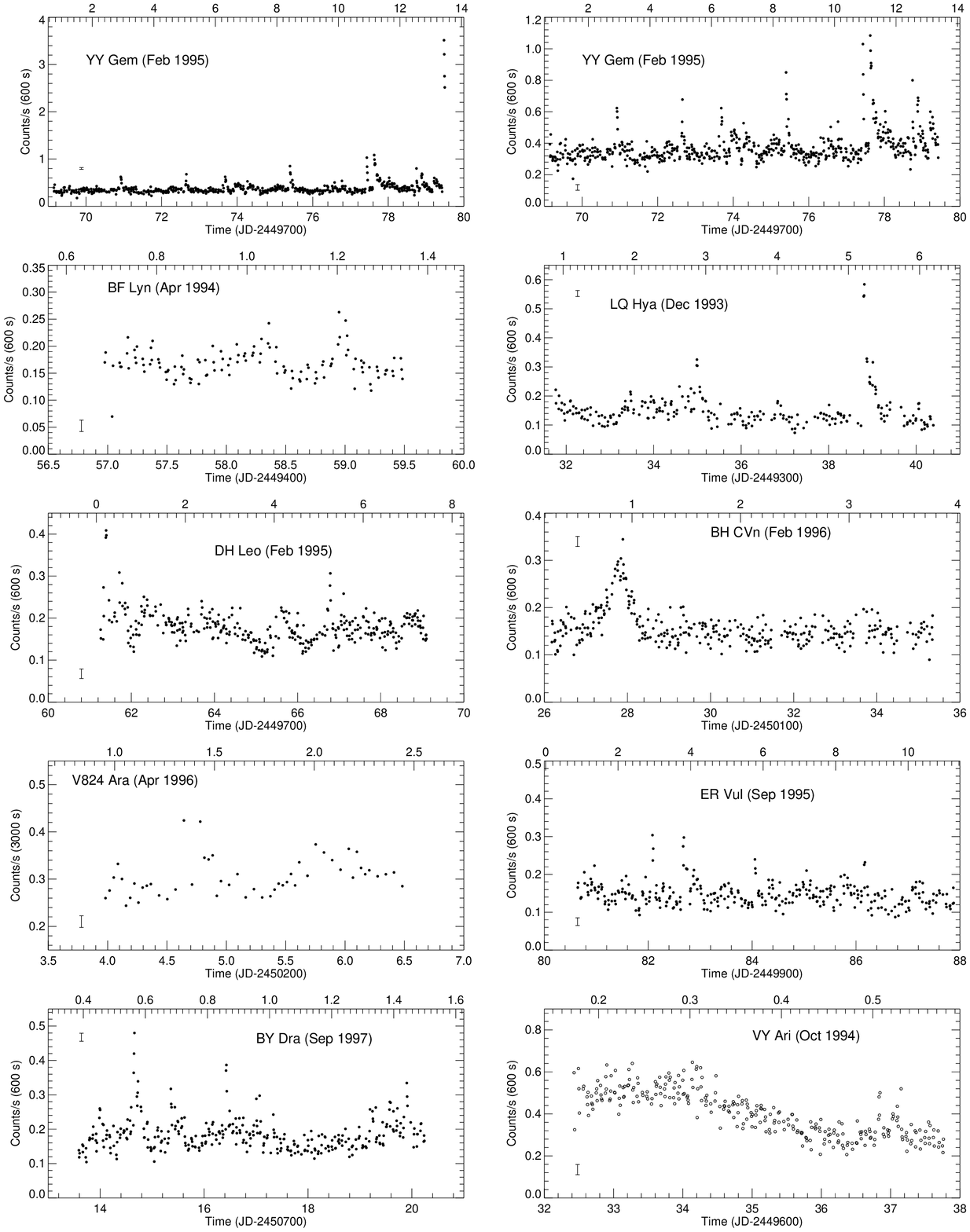}
  \caption{(c) continued.}
\end{figure}

\setcounter{figure}{0}
\begin{figure}
\plotone{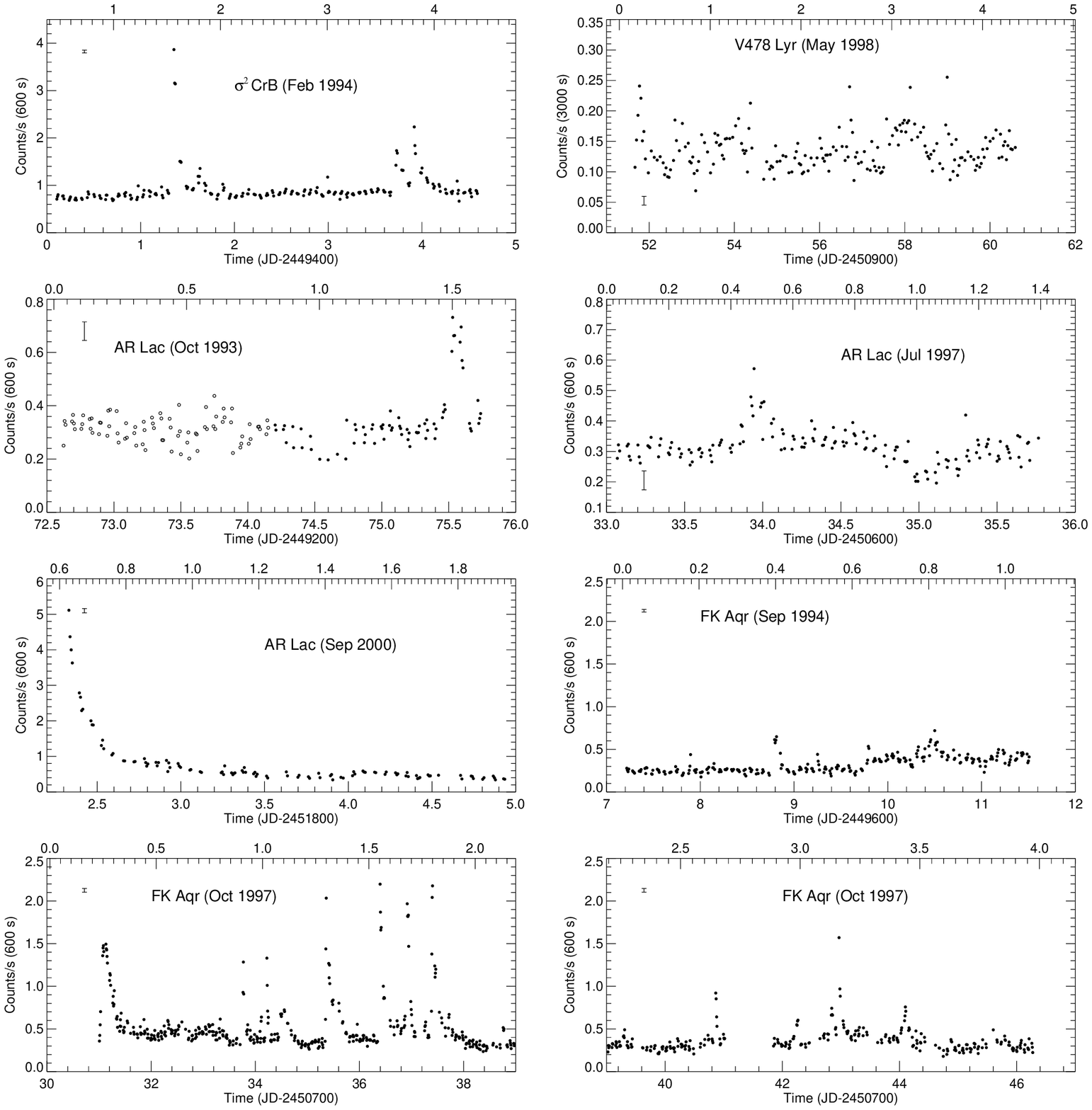}
  \caption{(d) continued.}
\end{figure}

\begin{figure}
  \epsscale{0.9}
\plotone{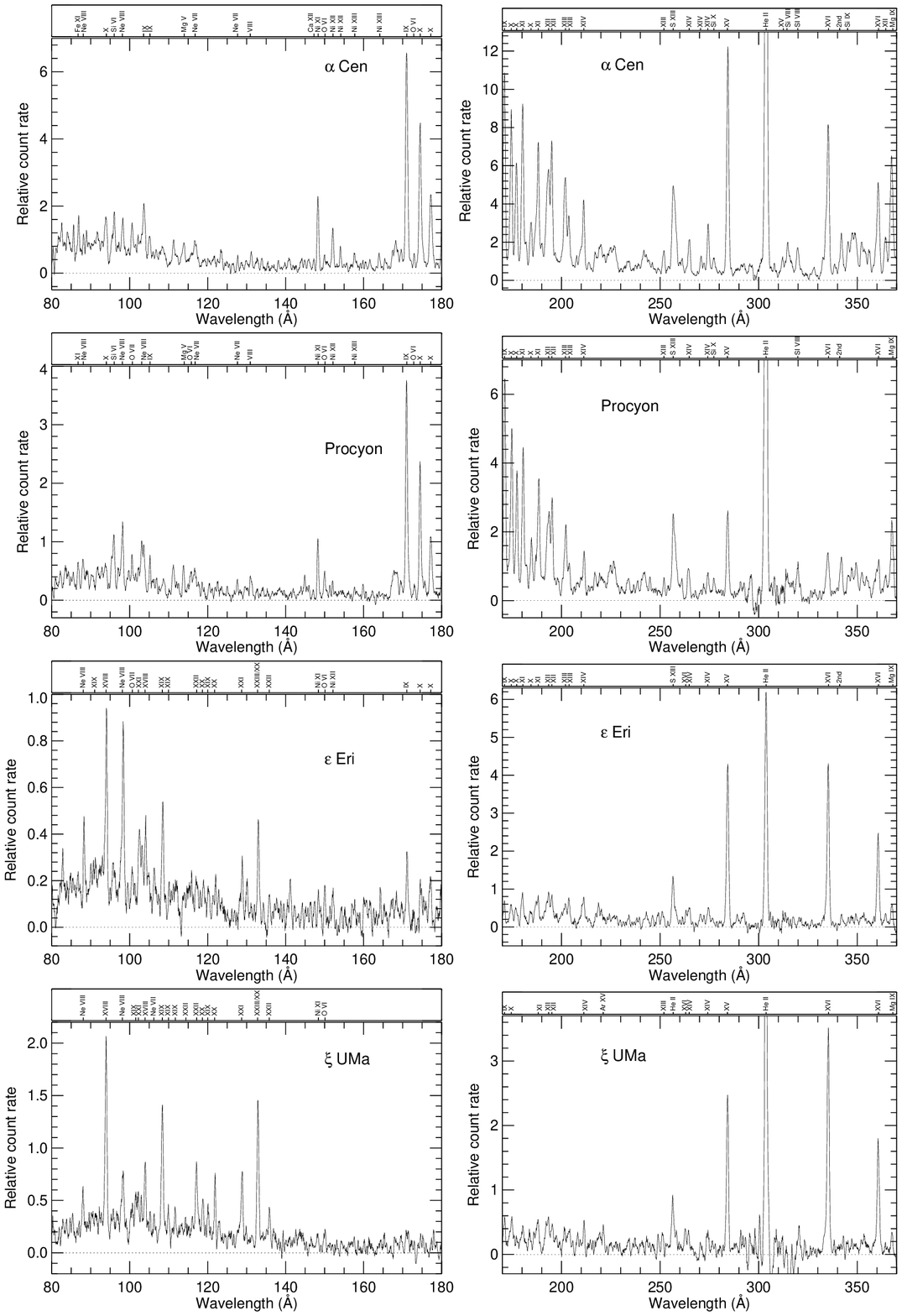}
  \caption{(a) EUVE SW and MW spectra of the stars in the sample. Ion
  stages of iron are marked in the top panel. Spectra are smoothed by
  5 pixels. Dotted lines indicate the zero flux level of each
  spectrum.  \label{specs}}
\end{figure}

\setcounter{figure}{1}
\begin{figure}
  \epsscale{0.9}
\plotone{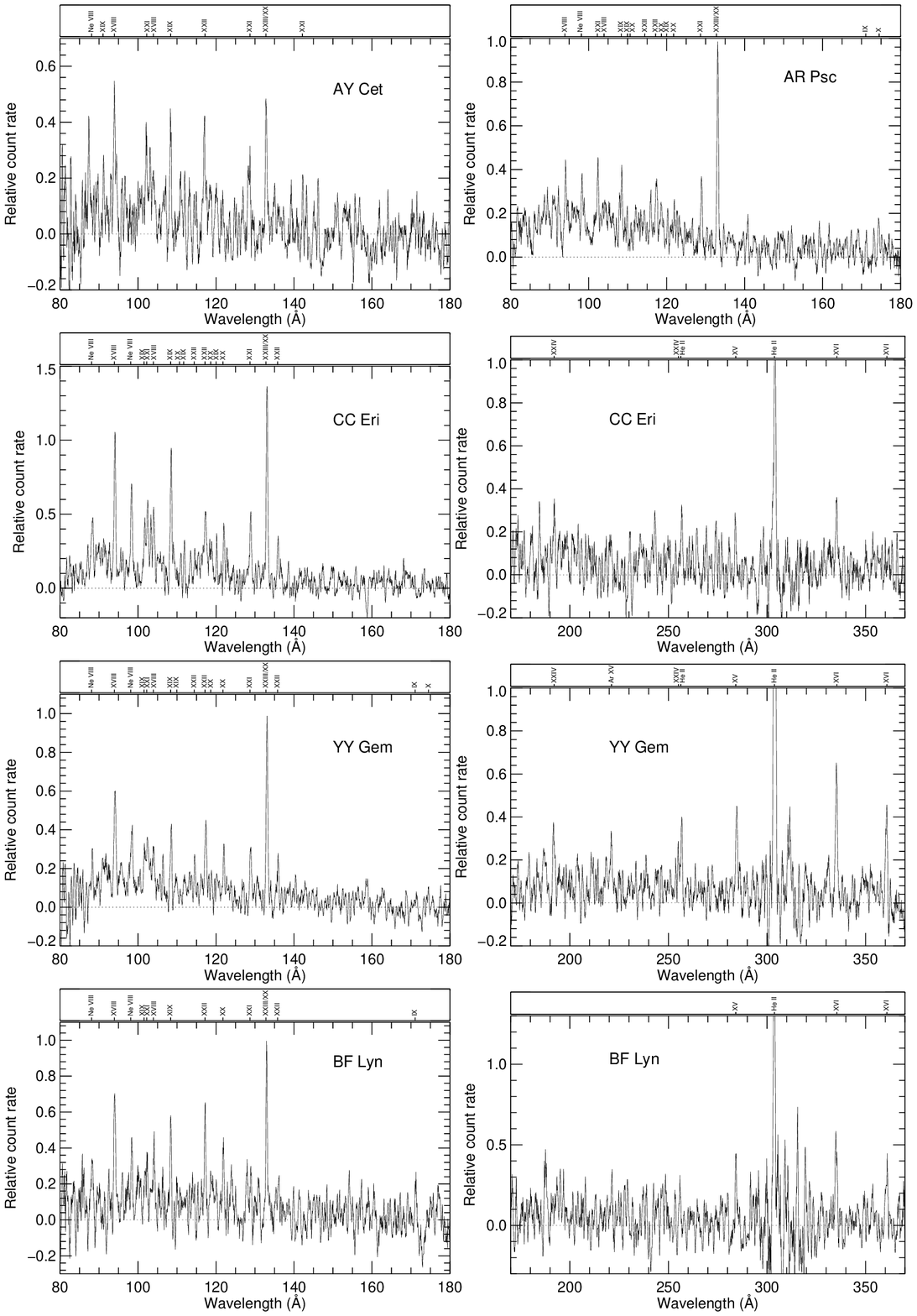}
  \caption{(b) continued.}
\end{figure}

\setcounter{figure}{1}
\begin{figure}
  \epsscale{0.9}
\plotone{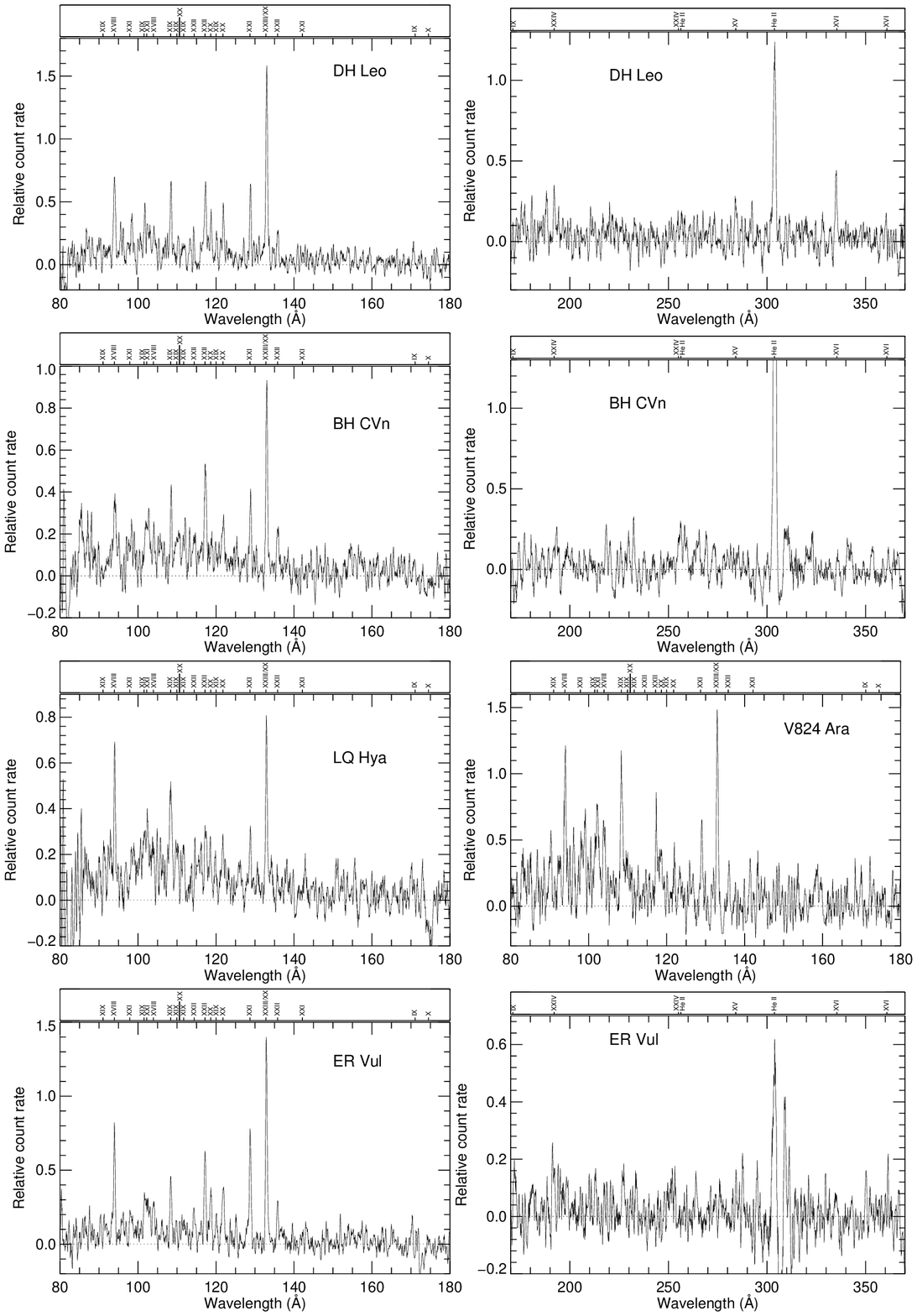}
  \caption{(c) continued.}
\end{figure}

\setcounter{figure}{1}
\begin{figure}
  \epsscale{0.9}
\plotone{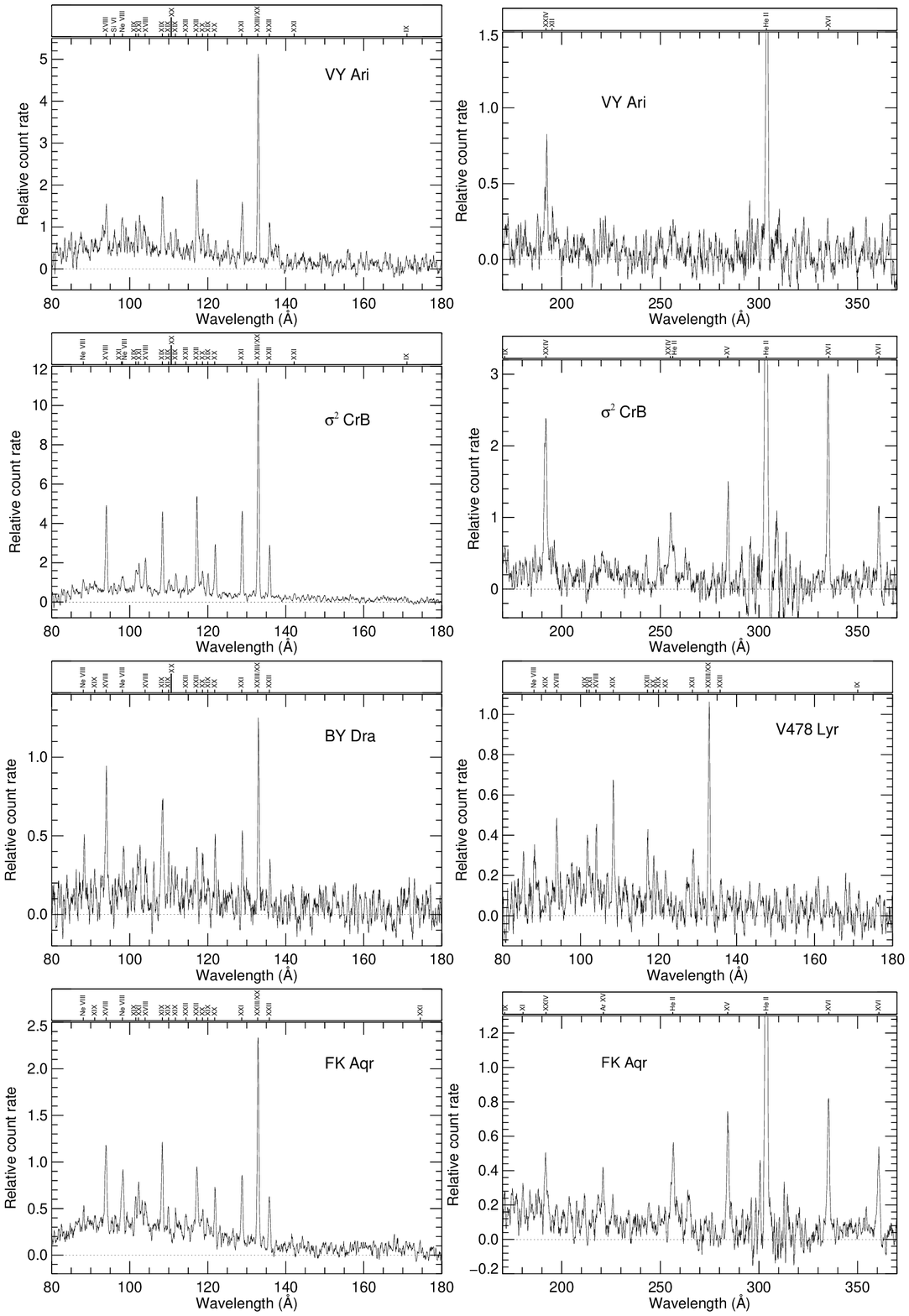}
  \caption{(d) continued.}
\end{figure}

\begin{figure}
  \epsscale{0.9}
\plotone{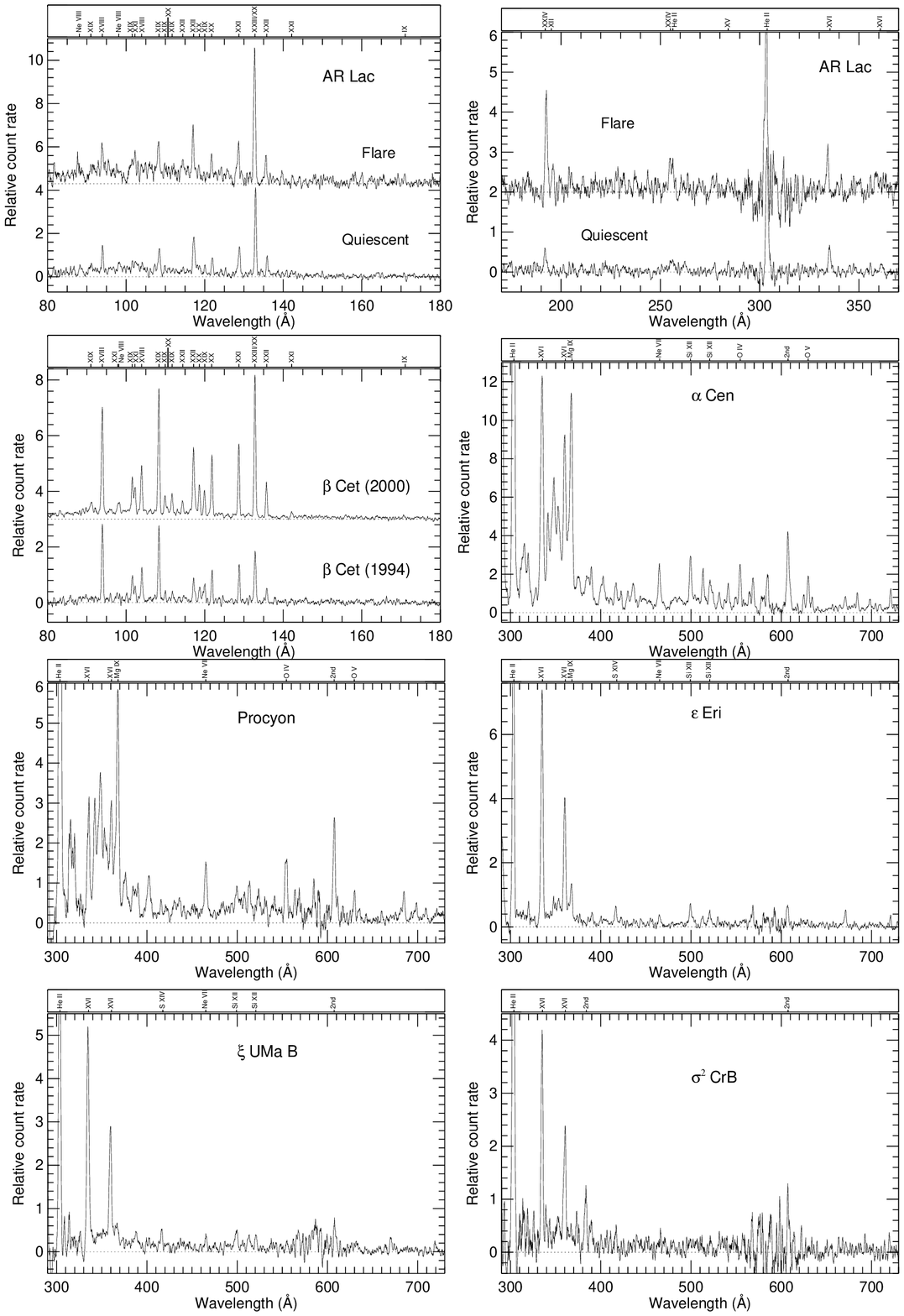}
  \caption{AR Lac, $\beta$~Cet and LW spectra of 5 of the stars. Note
    that 2$^{nd}$ order of the \ion{He}{1} line is detected in the LW
    spectra at 608~\AA.\label{specesp}}
\end{figure}

\clearpage

\begin{figure}
  \epsscale{1.}
\plotone{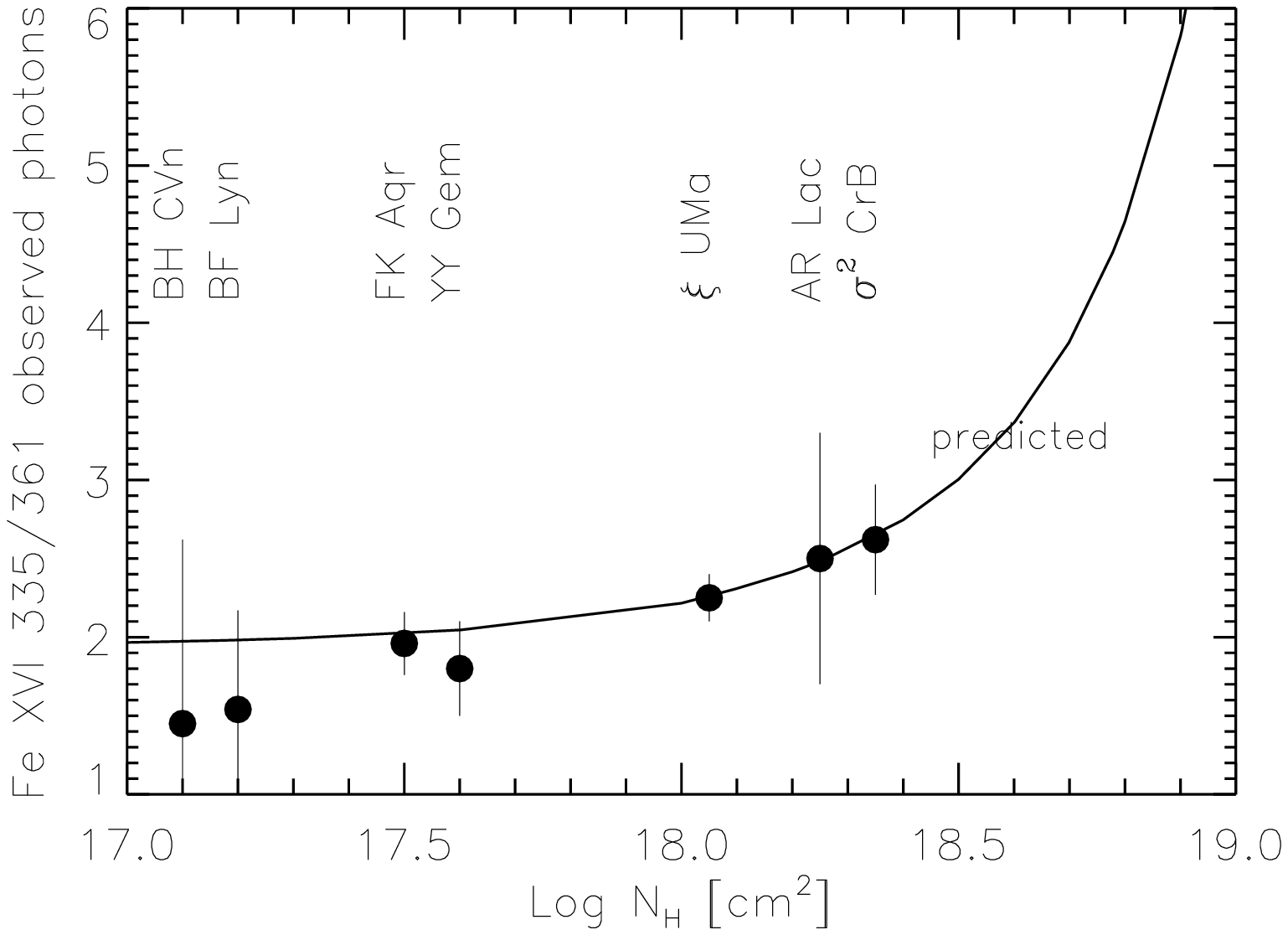}
  \caption{\ion{Fe}{16} \gl335 and \gl361 line flux ratios (in photon
  units) expected for different values of interstellar hydrogen column
  density (N$_H$), and observed ratios for several stars in the
  sample, with 1-$\sigma$ observational error bars. 
  \label{ismcalc}}
\end{figure}

\begin{figure}
\epsscale{0.80}
\plotone{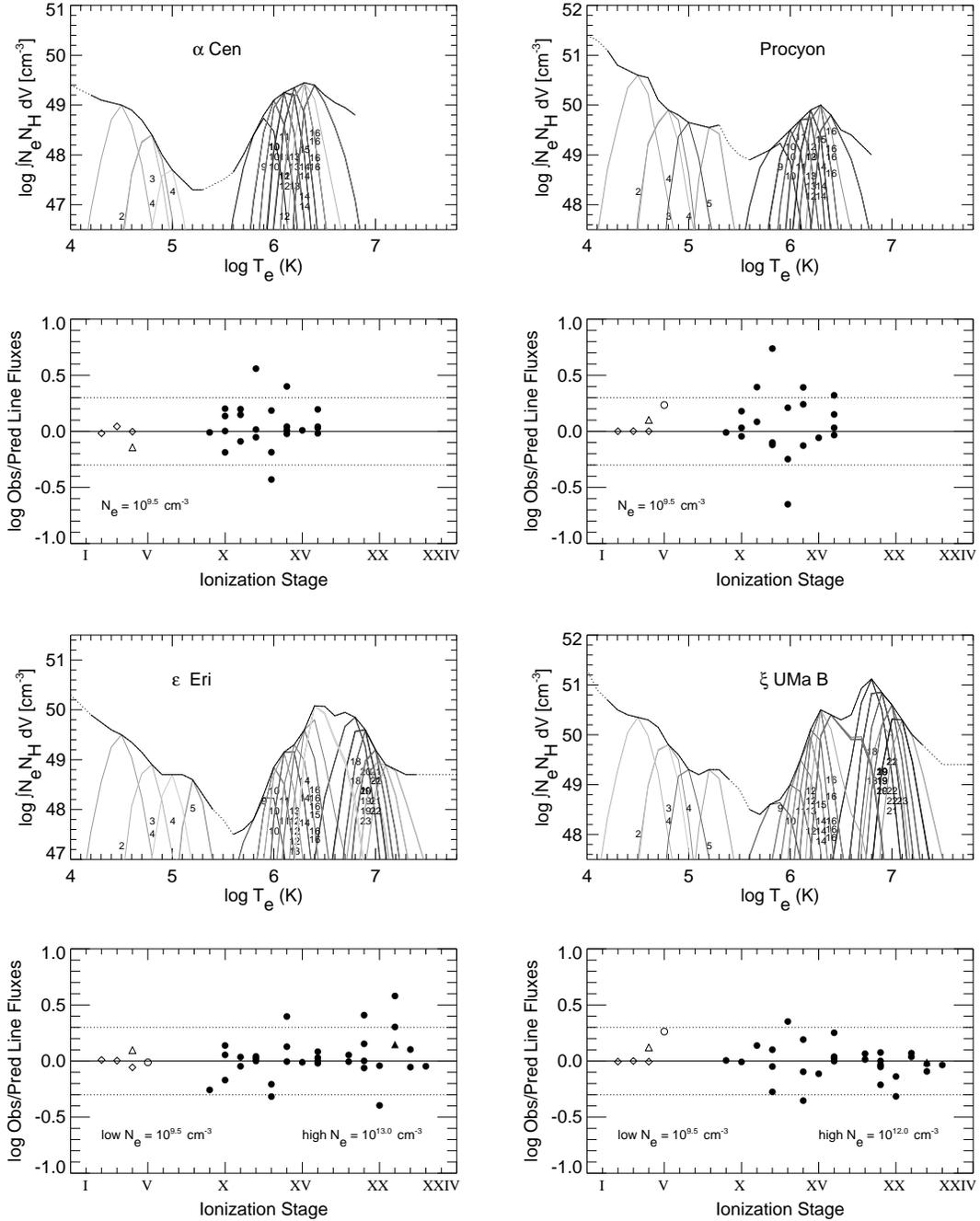}
  \caption{(a){\it Upper panels}: EMD for the summed EUVE 
spectrum combined with the IUE spectrum for each star. Thin lines 
represent the relative contribution function for each ion (the 
emissivity function multiplied by the EMD at each point).
{\it Lower panels}: Observed-to-predicted line ratios for the ion 
stages in top figure with S/N greater than 3. The dotted line 
denotes a factor of 2. Symbols used are open circles for N, diamonds
for C, and open triangles for Si. Fe lines with S/N
higher than 4 are denoted with filled circles, solid triangles for
those with S/N between 3 and 4, and the plus sign (+) for S/N less
than 3. Electron densities used in the calculation of the EMD are shown; if a value
is not given, then the density was taken as 10$^{12}$ cm$^{-3}$.
  \label{emdfigs}}
\end{figure}
\clearpage

\setcounter{figure}{4}
\begin{figure}
\plotone{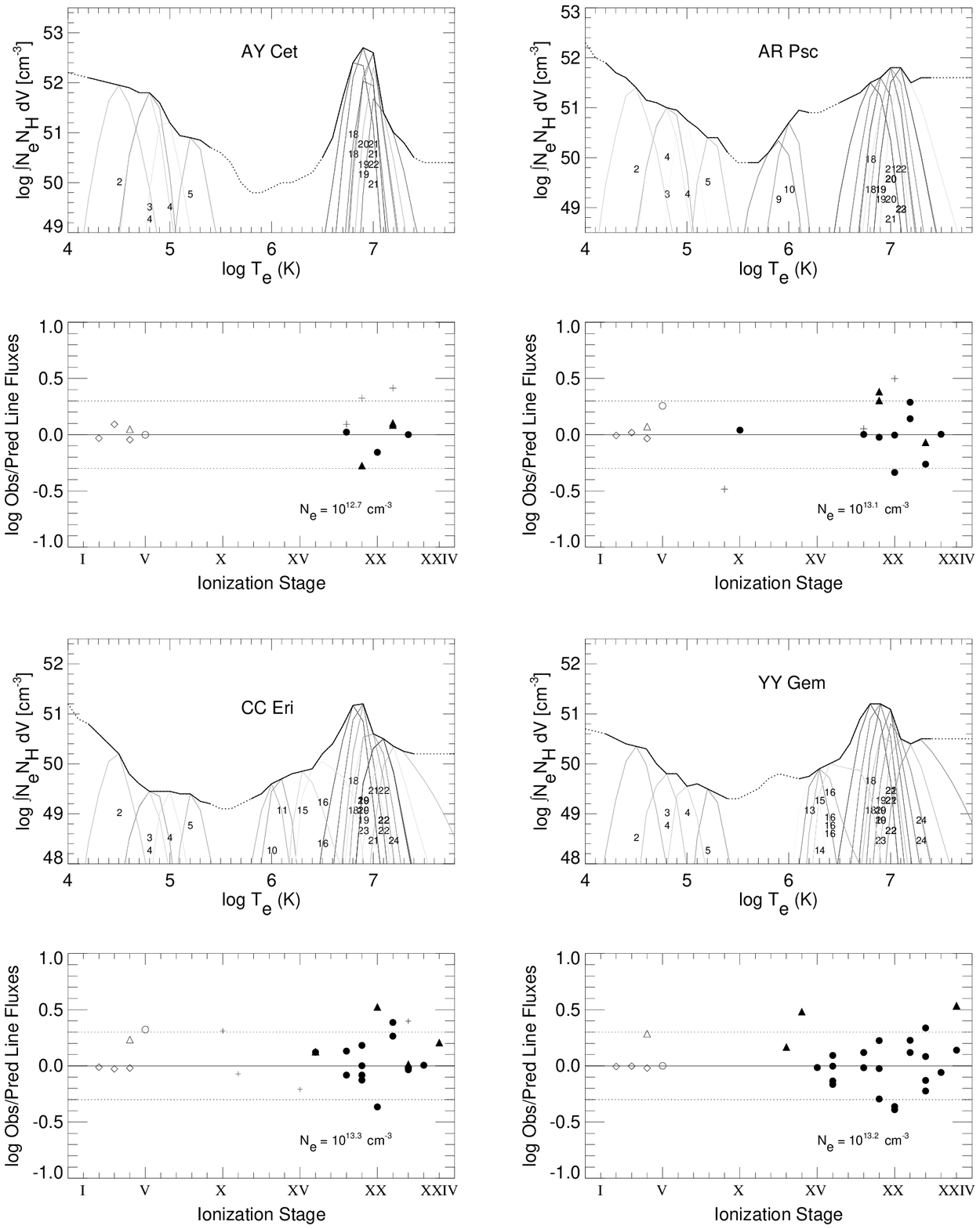}
  \caption{(b) continued.} 
\end{figure}

\setcounter{figure}{4}
\begin{figure}
\plotone{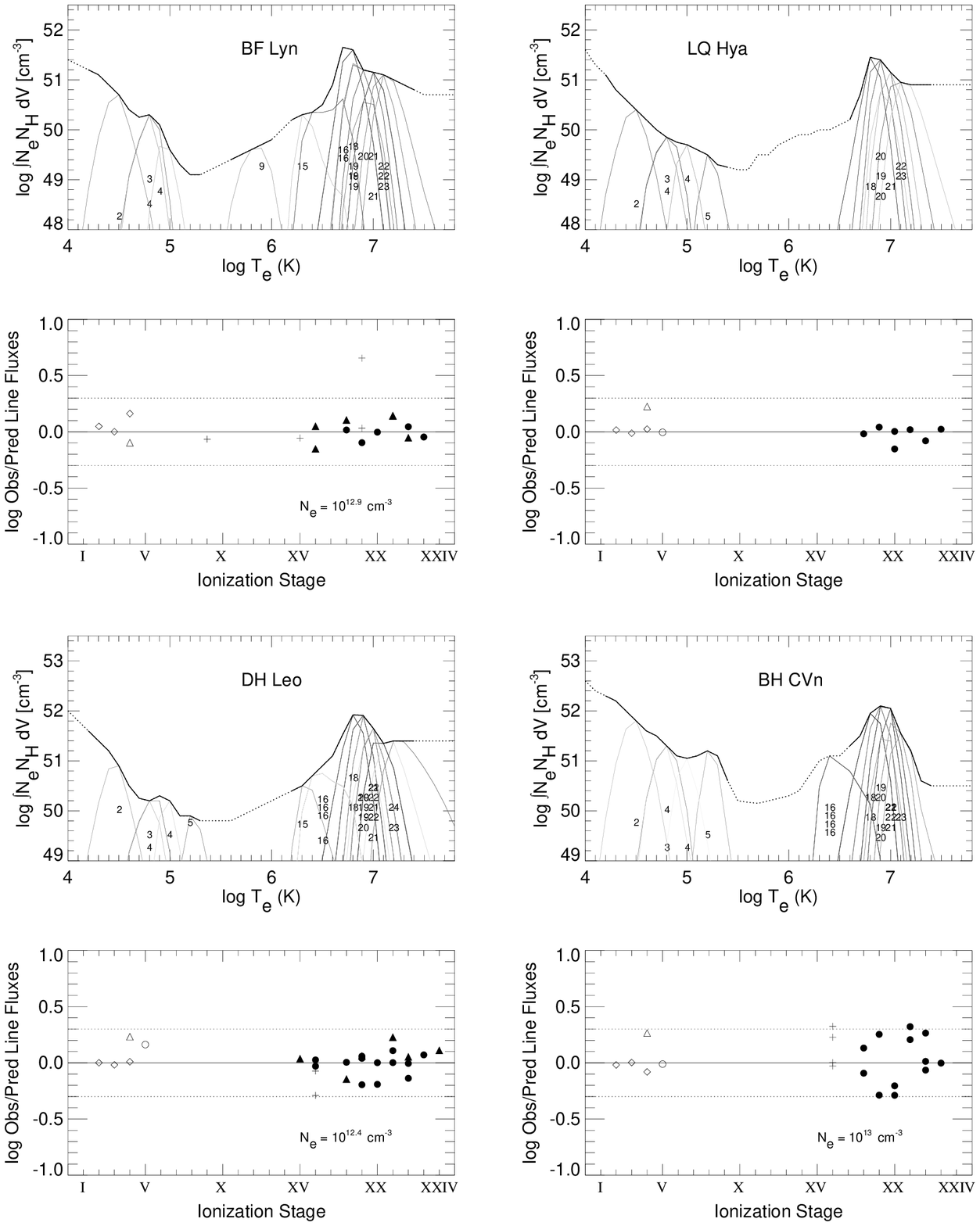}
  \caption{(c) continued.} 
\end{figure}

\setcounter{figure}{4}
\begin{figure}
\plotone{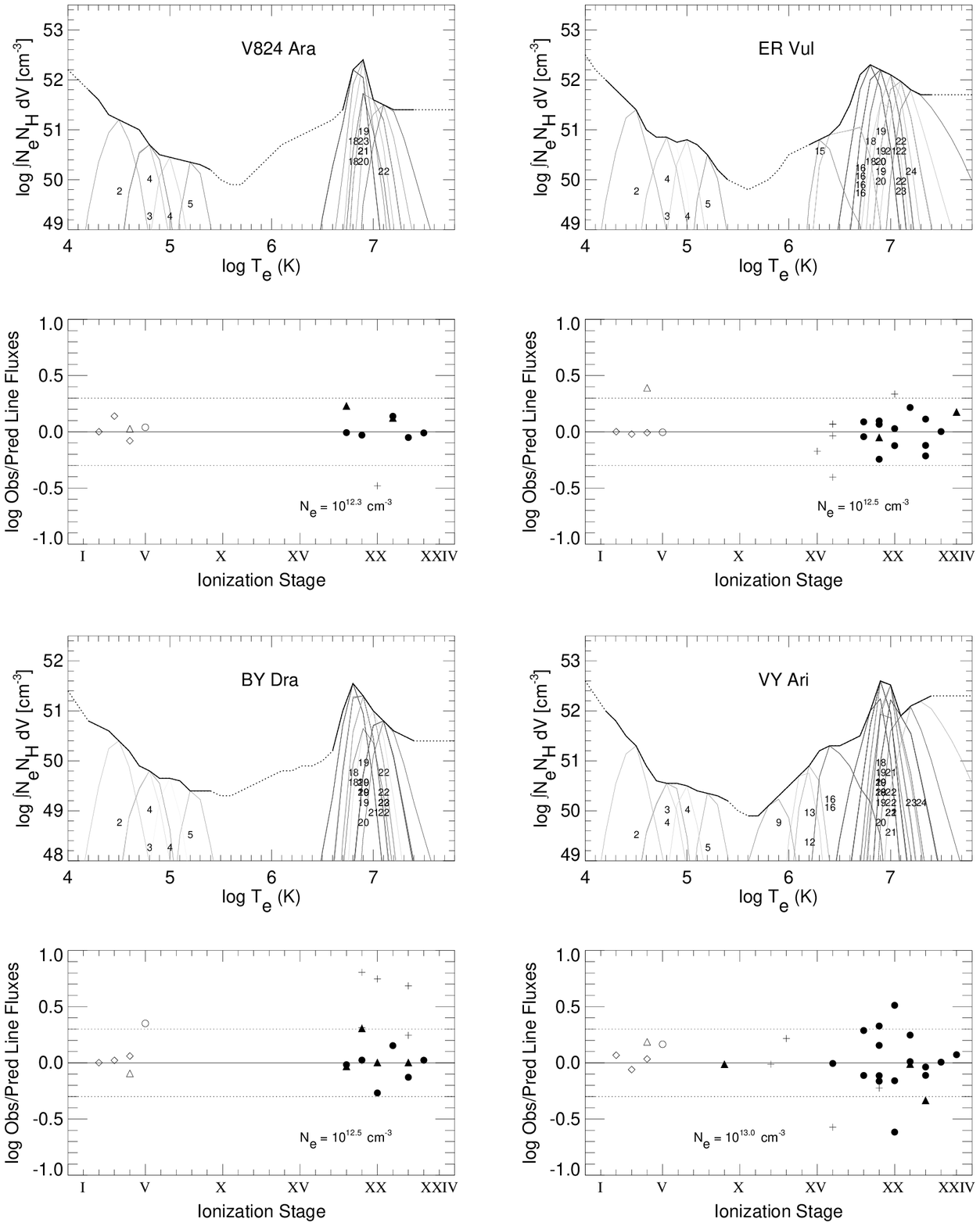}
  \caption{(d) continued.} 
\end{figure}

\setcounter{figure}{4}
\begin{figure}
\plotone{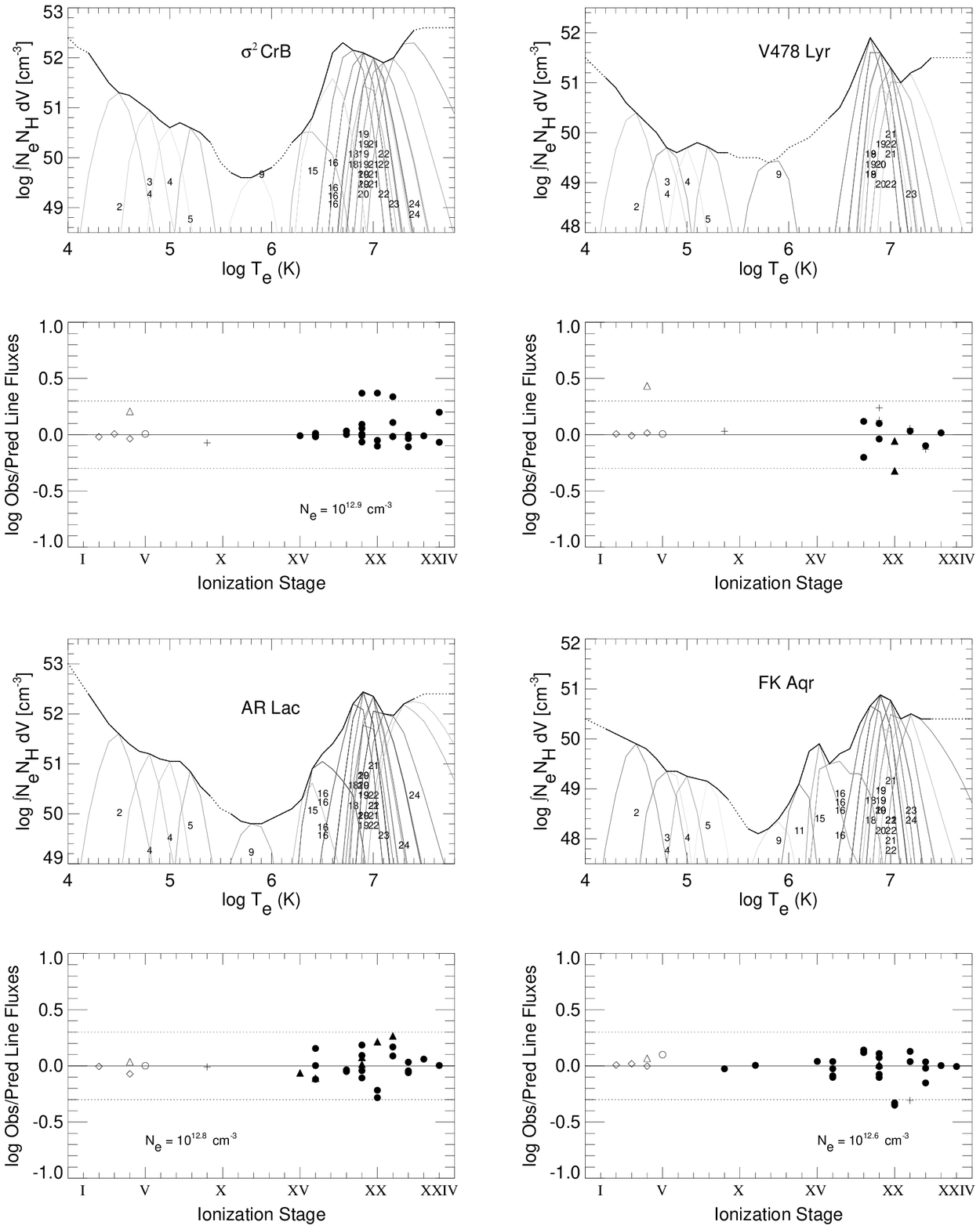}
  \caption{(e) continued.} 
\end{figure}

\clearpage

\setcounter{figure}{4}
\begin{figure}
  \epsscale{0.5}
\plotone{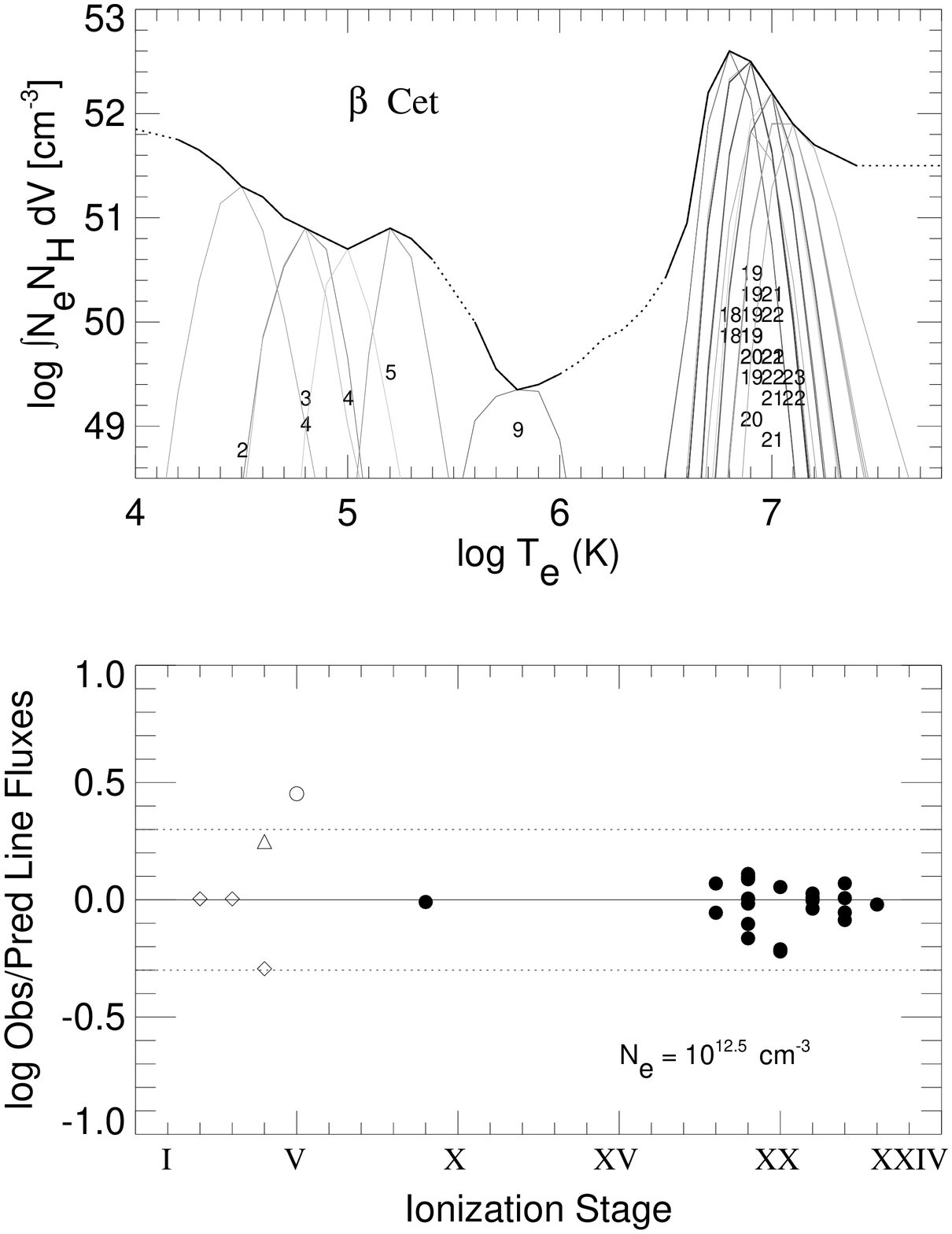}
  \caption{(f) continued.} 
  \epsscale{1.}
\end{figure}

\begin{figure}
\epsscale{1.0}
\plotone{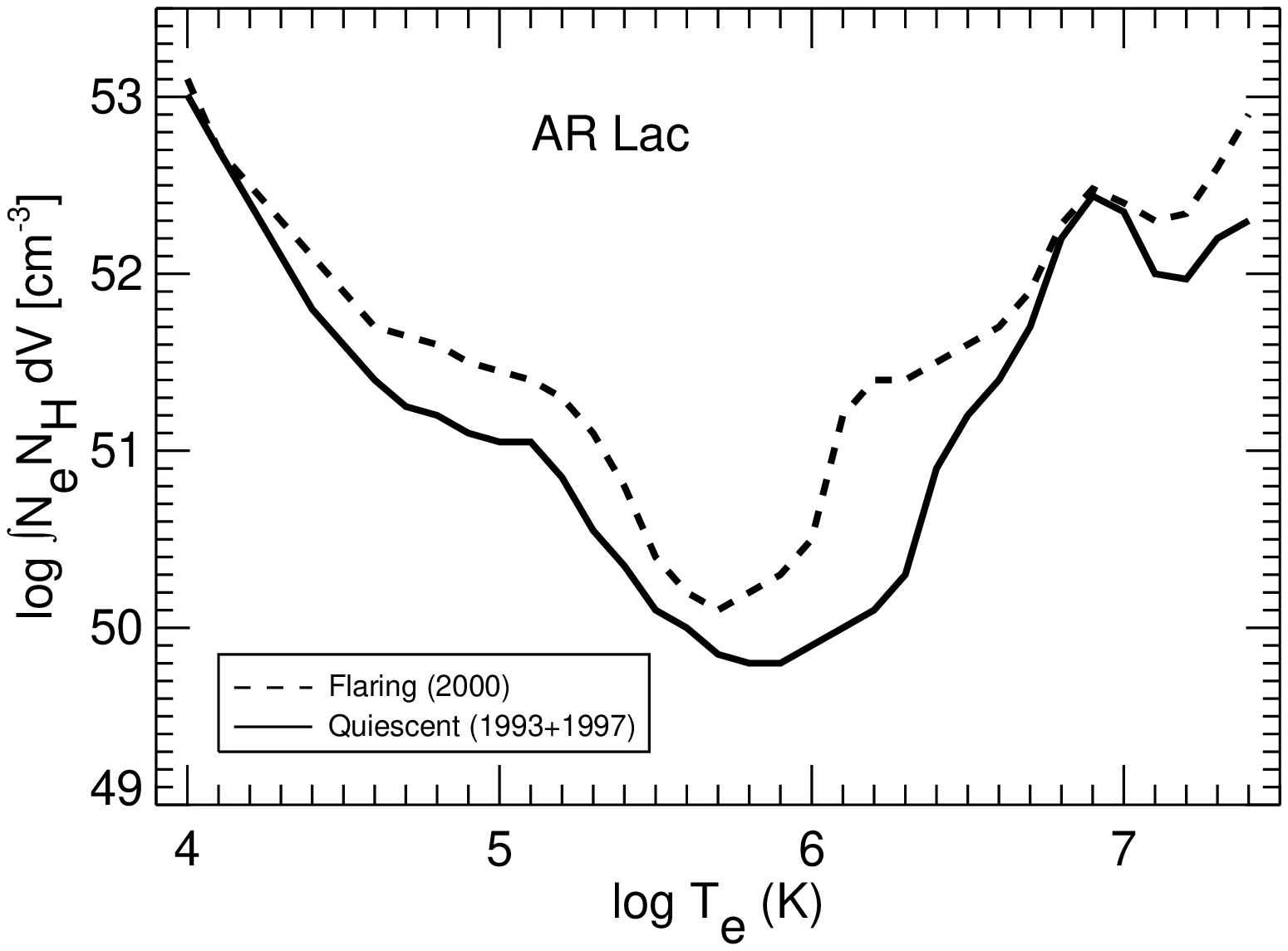}
  \caption{Comparison between the EMD calculated from the observations
    on AR~Lac in 1993+1997 and those in 2000.} 
  \label{stagesarlac}
\end{figure}

\begin{figure}
\plotone{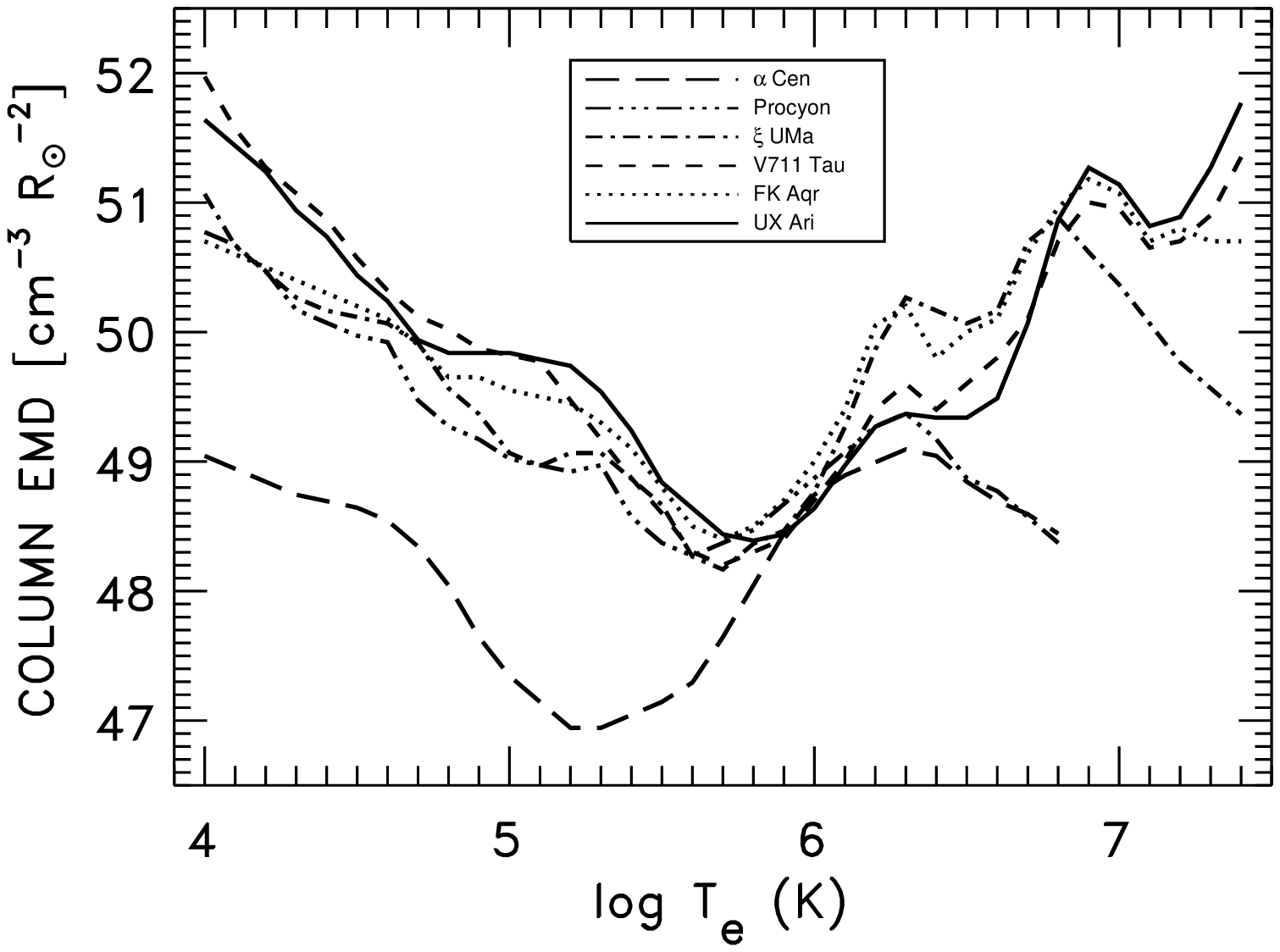}
  \caption{EMD of selected stars normalized to the solar photospheric
    radius [EMD/$4 \pi (R_1^2+R_2^2)$= ``column'' EMD, 
    assuming that both stars in binary systems 
    contribute to the observed emission, with radii measured in
    R$_\odot$]}
    \label{sevstars}
\end{figure}

\begin{figure}
\plotone{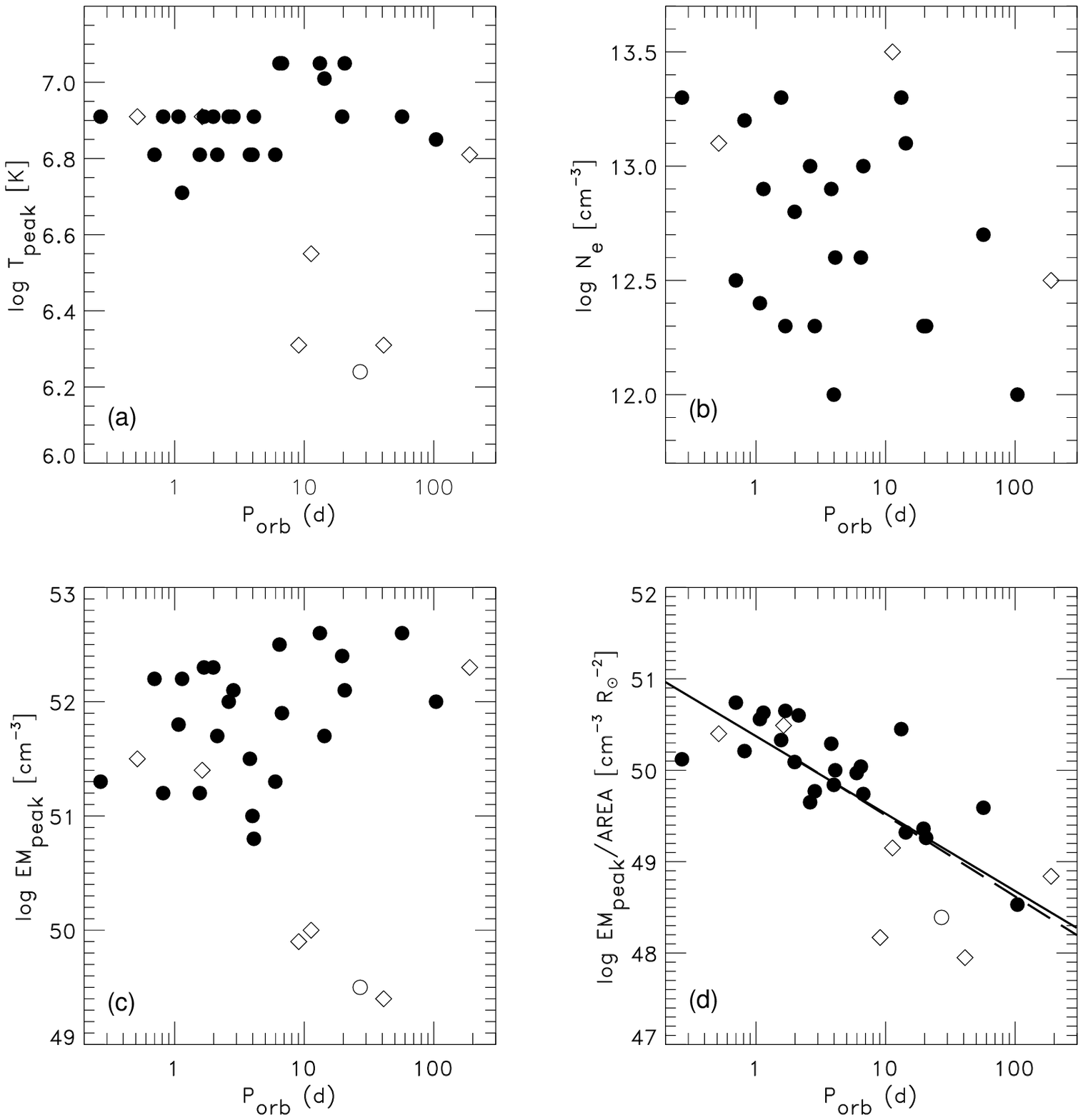}
  \caption{Orbital (or photometric) period relations with several features for 30
    single and binary
    stars: (a) Temperature at the peak [3 largest
    values of the emission measure (EM)], (b) density at
    log~T(K)$\sim$6.9, (c) EM at 
    the peak, (d) EM at the peak per unit area [EM/$4 * \pi
  *(R_1^2+R_2^2)$], with radii in solar units. A solid line with the
    best fit to all the data has been plotted in (d), along with a dashed
    line representing the fit including only objects with period over
    1 day.  Filled circles represent binary systems, diamonds are
    single stars, and an open circle represents the Sun during the
    solar maximum \citep[from][]{orl00}.} 
    \label{periods}
\end{figure}

\begin{figure}
  \epsscale{0.5}
\plotone{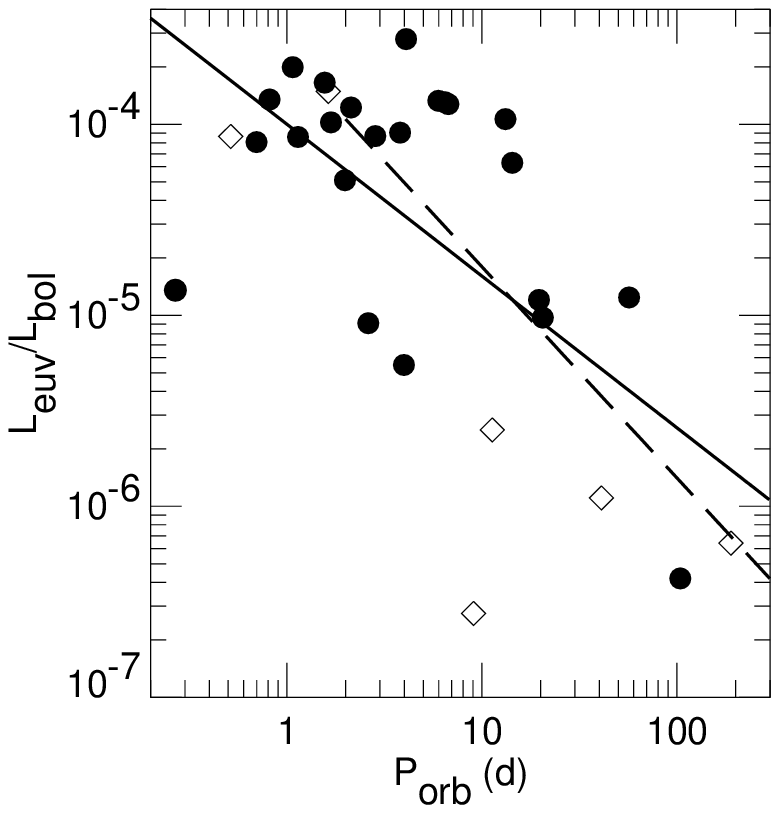}
  \caption{Orbital period versus EUV (80--170~\AA) luminosity
    weighted by the bolometric luminosity (see text). A solid line with the
    best fit to all the data has been plotted in (d), along with a dashed
    line representing the fit including only objects with period
  longer than 
    2 days.  Filled circles represent binary systems and diamonds are
    single stars.} 
    \label{lumperiod}
  \epsscale{1.}
\end{figure}

\clearpage

\begin{deluxetable}{lrcrrrrrr}
  \tabletypesize{\footnotesize}
  \tablewidth{0pt}
  \tablecolumns{9}
  \tablecaption{Stellar parameters\label{tabparam}}
  \tablehead{\colhead{Name}  & \colhead{HD} & \colhead{Spectral Type}
  & \colhead{P$_{orb}$(d)} & \colhead{P$_{phot}$(d)} 
    & \colhead{i ($\arcdeg$)} & \colhead{N$_H$(cm$^{-2}$)} &
  \colhead{d (pc)\tablenotemark{a}} & \colhead{R$_\star$(R$_\odot$)}
  }
  \startdata
  $\beta$ Cet & 4128 & K0III &\nodata& \nodata& 60: & 2.2$\times 10^{18}$ & 
  29.4 & 15.1:\\
  {AY Cet} & {7672} & {WD/G5III} & {56.824}& {77.22}&
  {29}\hspace{1pt}\tablenotemark{c}& {6.$\times 10^{18}$}\hspace{1pt}\tablenotemark{d} & {78.5} &
  {0.12/9}\hspace{1pt}\tablenotemark{c} \\ 
  {AR Psc} & {8357} & {G7V/K1IV} & {14.3023}\hspace{1pt}\tablenotemark{f}&
  {12.38}\hspace{1pt}\tablenotemark{e}& 
  {30:}\hspace{1pt}\tablenotemark{f} & {2.$\times 10^{18}$}\hspace{1pt}\tablenotemark{g} & {45.2} &
  {?/3.2}\hspace{1pt}\tablenotemark{f} \\
  {CC Eri} & {16157} & {K7V/M3V}\hspace{1pt}\tablenotemark{h} & {1.56145}&
  {1.56145}\hspace{1pt}\tablenotemark{e}& 
  {42} & {2.6$\times 10^{18}$}\hspace{1pt}\tablenotemark{i} &  {11.5} &
  {0.65/0.41}\hspace{1pt}\tablenotemark{j} \\ 
  {VY Ari} & {17433} & {K3-4V-IV+?} & {13.198}& {16.23}\hspace{1pt}\tablenotemark{k}&
  {60:} & {1.5$\times 10^{18}$}\hspace{1pt}\tablenotemark{l} &  {44.0} &
  {$\ga$3.3/?}\hspace{1pt}\tablenotemark{m}\\ 
  {$\epsilon$ Eri} & {22049} & {K2V} & \nodata& {11.3}\hspace{1pt}\tablenotemark{n}&
  {30:}\hspace{1pt}\tablenotemark{p} & {1.3$\times 10^{18}$}\hspace{1pt}\tablenotemark{g} &
  {3.22} & {0.75:}\hspace{1pt}\tablenotemark{l} \\ 
  {YY Gem} & {60179} & {dM1e/dM1e} & {0.814282}\hspace{1pt}\tablenotemark{q}& {0.8143}&
  {86.3}\hspace{1pt}\tablenotemark{q} & {6.$\times 10^{17}$}\hspace{1pt}\tablenotemark{l} &
  {15.8} & {0.62/0.62}\hspace{1pt}\tablenotemark{q} \\
  {Procyon} & {61421} & {F5IV-V} & \nodata& \nodata&
  {32}\hspace{1pt}\tablenotemark{r} & {1.6$\times 10^{18}$} & {3.50} &
  {2.06}\hspace{1pt}\tablenotemark{r} \\
\enddata
\tablecomments{Reference (b) was used when no other reference is
indicated; data for $\beta$~Cet as in \citet{paper1}.}
\tablerefs{(a) \citet{hippa}; 
    (b) \citet{cabs}; (c) \citet{sch95}; (d) \citet{diam95};
    (e) \citet{cuti01}; (f) \citet{fek96}; (g) based on \citet{fru94};
    (h) \citet{cuti98}; (i) \citet{pan95}; (j) \citet{ama00}; 
    (k) \citet*{str99}; (l) Present work; (m) \citet{fek97}; 
    (n) \citet{bali83}; (p) \citet{saa97}; (q)
    \citet{torres01}; (r) \citet{irw92}; (s) \citet{bar86}; (t)
    \citet{don99}; (u) \citet{griffin98}; (v) \citet{lin96}; (w)
    \citet{mor00}; (x) \citet{pas91}; (y) \citet{ruc98}; (z)
\citet{mar98}}
\end{deluxetable}


%
\begin{deluxetable}{llrrr}
\tablewidth{0pt}
\tablecolumns{5}
\tablecaption{Exposure Times for EUVE Spectrographs \label{euvetimes}}
\tablehead{
    \colhead{}& \colhead{}& \multicolumn{3}{c}{Exposure time (ks)} \\ 
    \cline{3-5}
    \colhead{Name} &\colhead{Start date}  & \colhead{SW} &
\colhead{MW} & \colhead{LW}     }
\startdata
    $\beta$ Cet     & {2000 Aug  5} & 808   & \nodata & \nodata \\
    AY Cet          & {1993 Sep 28} & 116   & \nodata & \nodata \\
    {AR Psc}        & {1997 Aug 26} & 388   & \nodata & \nodata \\
    {CC Eri}        & {1995 Sep 13} & {257} & 144 & 132 \\
    {VY Ari}        & {1994 Oct 6}  & {244} & 159 & 156 \\
    {$\epsilon$ Eri}& {1993 Oct 22} & { 87} &  62 &  60 \\
    {$\epsilon$ Eri}& {1995 Aug 31} & {495} & 240 & 214 \\
    {Capella}       & {2001 Jan 14} & 49    &  48 &  48 \\
\enddata
\end{deluxetable}


\begin{deluxetable}{lrrcrcrcrcrcrcrcrcrcrcrc}
\tabletypesize{\footnotesize}              
\tabcolsep 2.2pt                           
\rotate
\tablewidth{0pt}                           
\tablecaption{EUVE line fluxes (I)\label{euveflux}}
\tablehead{
 \colhead{}& \colhead{$\lambda$$_{lab}$} & \multicolumn{2}{c}{AY Cet}
 & \multicolumn{2}{c}{$\epsilon$ Eri} & \multicolumn{2}{c}{Procyon} &
 \multicolumn{2}{c}{BF Lyn} & \multicolumn{2}{c}{$\xi$ UMa} &
 \multicolumn{2}{c}{$\alpha$ Cen} & \multicolumn{2}{c}{V824 Ara} &
 \multicolumn{2}{c}{LQ Hya} & \multicolumn{2}{c}{BY Dra} &
 \multicolumn{2}{c}{BH CVn} & \multicolumn{2}{c}{$\beta$ Cet (2000)} \\ 
\colhead{Ion} & \colhead{(\AA)} & \colhead{S/N} & \colhead{Flux} &
\colhead{S/N} & \colhead{Flux} & \colhead{S/N} & \colhead{Flux} &
\colhead{S/N} & \colhead{Flux} & \colhead{S/N} & \colhead{Flux} &
\colhead{S/N} & \colhead{Flux} & \colhead{S/N} & \colhead{Flux} &
\colhead{S/N} & \colhead{Flux} & \colhead{S/N} & \colhead{Flux} &
\colhead{S/N} & \colhead{Flux} & \colhead{S/N} & \colhead{Flux}
}
\startdata
\\
\multicolumn{22}{l}{\em Short Wavelength Spectrometer} \\
\\
\ion{Fe}{11} &  86.77 & \nodata & \nodata & \nodata & \nodata & $^* 11.5$ & 2.17e-04 & \nodata & \nodata & \nodata & \nodata & $^* 11.1$ & 4.35e-04 & \nodata & \nodata & \nodata & \nodata & \nodata & \nodata & \nodata & \nodata & \nodata & \nodata\\
\ion{Ne}{ 8}\tablenotemark{a} &  88.08 &   3.6 & 1.54e-04 &   8.3 & 1.72e-04 &  10.2 & 2.56e-04 &   3.9 & 1.98e-04 &   8.0 & 1.91e-04 &   6.8 & 1.98e-04 & \nodata & \nodata & \nodata & \nodata &   3.7 & 1.69e-04 &   5.3 & 9.75e-05 & \nodata & \nodata\\
\ion{Fe}{19}\tablenotemark{b} &  91.02 &   2.6 & 8.14e-05 &   5.3 & 7.49e-05 & \nodata & \nodata & \nodata & \nodata & \nodata & \nodata & \nodata & \nodata & \nodata & \nodata & \nodata & \nodata &   2.5 & 9.56e-05 & \nodata & \nodata & 15.1 & 2.74e-04 \\
\ion{Fe}{18} &  93.92 &   4.9 & 2.11e-04 &  16.0 & 3.68e-04 & \nodata & \nodata &  6.7 & 2.91e-04 &  21.6 & 8.25e-04 &  \nodata & \nodata &  5.1 & 5.13e-04 &   7.8 & 2.26e-04 &   7.4 & 3.59e-04 &   8.3 & 1.75e-04 & 44.7 & 1.44e-03 \\
\ion{Fe}{10} &  94.01 & \nodata & \nodata & \nodata & \nodata & $^* 10.2$ & 2.42e-04 &  \nodata & \nodata & \nodata & \nodata & $^* 14.9$ & 5.70e-04 &\nodata & \nodata & \nodata & \nodata & \nodata & \nodata & \nodata & \nodata & \nodata\\
\enddata
\end{deluxetable}

\newpage

\begin{deluxetable}{lrrcrcrcrcrcrcrcrcrcrcrc}
\tabletypesize{\footnotesize}              
\tabcolsep 2.2pt                           
\rotate
\tablewidth{0pt}
\tablecaption{EUVE line fluxes (II)\label{euveflux2}}
\tablehead{
 \colhead{}& \colhead{$\lambda$$_{lab}$} & \multicolumn{2}{c}{$\sigma ^2$ CrB} & \multicolumn{2}{c}{AR Psc} & \multicolumn{2}{c}{CC Eri} & \multicolumn{2}{c}{VY Ari} & \multicolumn{2}{c}{V478 Lyr} & \multicolumn{2}{c}{AR Lac (q)} & \multicolumn{2}{c}{AR Lac (f)} & \multicolumn{2}{c}{YY Gem} & \multicolumn{2}{c}{DH Leo} & \multicolumn{2}{c}{ER Vul} & \multicolumn{2}{c}{FK Aqr}  \\
\colhead{Ion} & \colhead{(\AA)} & \colhead{S/N} & \colhead{Flux} & \colhead{S/N} & \colhead{Flux} & \colhead{S/N} & \colhead{Flux} & \colhead{S/N} & \colhead{Flux} & \colhead{S/N} & \colhead{Flux} & \colhead{S/N} & \colhead{Flux} & \colhead{S/N} & \colhead{Flux} & \colhead{S/N} & \colhead{Flux} & \colhead{S/N} & \colhead{Flux} & \colhead{S/N} & \colhead{Flux} & \colhead{S/N} & \colhead{Flux}
}
\startdata
\multicolumn{22}{l}{\em Short Wavelength Spectrometer} \\
\ion{Ne}{ 8}\tablenotemark{a} &  88.08 &  22.0 & 2.76e-04 & \nodata & \nodata &   6.3 & 2.14e-04 & \nodata & \nodata &   4.1 & 1.53e-04 & \nodata & \nodata & 3.2 & 3.42e-04 &   4.6 & 1.08e-04 &   2.5 & 6.01e-05 & \nodata & \nodata &   4.6 & 8.66e-05 \\
\ion{Fe}{19}\tablenotemark{a} &  91.02 & \nodata & \nodata & \nodata & \nodata & \nodata & \nodata & \nodata & \nodata &   2.2 & 5.89e-05 &   4.3 & 1.41e-04 & 2.0 & 1.61e-04 & \nodata & \nodata & \nodata & \nodata & \nodata & \nodata &   3.9 & 6.27e-05 \\
\ion{Fe}{18} &  93.92 &  22.8 & 1.72e-03 &   5.0 & 9.11e-05 &  10.1 & 3.62e-04 &   9.5 & 3.79e-04 &   5.1 & 1.84e-04 &   9.6 & 4.46e-04 & 5.6 & 6.28e-04 &  10.4 & 2.71e-04 &   9.6 & 3.20e-04 &  10.0 & 2.90e-04 &  16.1 & 4.46e-04 \\
\ion{Si}{ 6} &  96.02 &   4.5 & 1.66e-04 & \nodata & \nodata & \nodata & \nodata &   5.1 & 1.79e-04 & \nodata & \nodata & \nodata & \nodata & \nodata & \nodata & \nodata & \nodata & \nodata & \nodata & \nodata & \nodata & \nodata & \nodata \\
\ion{Fe}{21} &  97.88 &   5.7 & 1.64e-04 & \nodata & \nodata & \nodata & \nodata & \nodata & \nodata & \nodata & \nodata & \nodata & \nodata & \nodata & \nodata & \nodata & \nodata & \nodata & \nodata & \nodata & \nodata & \nodata & \nodata \\
\ion{Ne}{ 8}\tablenotemark{a} &  98.13 &   7.1 & 3.79e-04 &   4.6 & 7.53e-05 &   7.8 & 2.27e-04 &   8.1 & 3.27e-04 & \nodata & \nodata &   6.6 & 2.59e-04 & 2.2 & 1.61e-04 &   7.8 & 1.65e-04 &   4.9 & 1.17e-04 &   5.1 & 1.05e-04 &  14.7 & 3.74e-04 \\
\ion{Fe}{19} & 101.55 &  10.5 & 4.63e-04 & \nodata & \nodata &   5.7 & 1.39e-04 &   5.6 & 1.80e-04 &   4.4 & 1.28e-04 &   5.3 & 1.78e-04 & 3.3 & 2.43e-04 &   6.0 & 1.07e-04 &   5.8 & 1.43e-04 &   9.7 & 2.79e-04 &   7.3 & 1.22e-04 \\
\enddata
\tablecomments{See description in Table~\ref{euveflux}. Also, AR~Lac (q) are the fluxes corresponding to the 1993 and 1997 summed observations, while AR~Lac (f) are the measurements of the 2000 campaign}
\tablenotetext{a}{See footnote in Table~\ref{euveflux} for the
  corresponding line describing blends.}
\tablenotetext{b}{May include blend of \ion{Fe}{12} \gl 192.39, \gl
  193.51, \ion{Fe}{11} \gl 192.83 and possibly other weaker components.}
\tablenotetext{c}{Nearby \ion{He}{2} emission ($\lambda$256.317) was 
deblended from the \ion{Fe}{24} ($\lambda$255.10) emission.}
\end{deluxetable}

\begin{deluxetable}{lrrlrrrrr}
\tabletypesize{\scriptsize}
\tabcolsep 5pt
\tablewidth{0pt}
\tablecaption{Line fluxes from IUE low resolution spectra.\label{tabiue}}
\tablehead{
\colhead{Star}  & \colhead{SWP} & \colhead{t (s)} & \colhead{Date} & 
\colhead{\ion{C}{3} $\lambda$1176} & \colhead{\ion{N}{5} $\lambda$1240} & 
\colhead{\ion{C}{2} $\lambda$1336} & \colhead{\ion{Si}{4} $\lambda$1396} & 
\colhead{\ion{C}{4} $\lambda$1549}}
\startdata
           AY Cet &  31246 &   5400 & 1987 Jun 25 & 5.00E-13 ( 7.9) & 6.67E-14 ( 5.0) & 3.60E-13 (14.9) & 2.42E-13 (10.4) & 9.06E-13 (30.5) \\
           AR Psc &  28524 &   7800 & 1986 Jun 21 & 2.47E-13 ( 7.3) & 1.26E-13 (11.6) & 3.57E-13 (23.1) & 1.45E-13 (17.1) & 7.58E-13 (45.6) \\
           CC Eri &  40462 &   5400 & 1990 Dec 12 & 1.18E-13 ( 4.1) & 1.91E-13 (20.4) & 3.40E-13 (30.4) & 1.10E-13 (14.3) & 5.37E-13 (24.7) \\
           VY Ari &  32099 &   3600 & 1987 Oct 15 & 8.89E-14 ( 3.4) & 8.88E-14 ( 7.8) & 3.82E-13 (32.3) & 8.00E-14 ( 9.9) & 4.70E-13 (26.9) \\
   $\epsilon$ Eri &  22032 &   4199 & 1984 Jan 18 & 4.35E-13 (11.1) & 1.59E-13 (11.2) & 8.49E-13 (42.3) & 2.93E-13 (23.6) & 1.17E-12 (48.3) \\
          Capella &  35279 &    600 & 1989 Jan 7 & 1.53E-11 (24.6) & 8.01E-12 (40.4) & 2.54E-11 (56.7) & 1.40E-11 (59.0) & 4.16E-11 (59.0) \\
           YY Gem &  18693 &   5700 & 1982 Nov 30 & 1.51E-13 ( 4.2) & 6.60E-14 ( 5.7) & 2.31E-13 (16.8) & 1.55E-13 (11.5) & 4.68E-13 (20.0) \\
          Procyon &  27836 &     60 & 1986 Mar  3 & 4.10E-12 ( 5.3) & 2.69E-12 ( 9.7) & 9.09E-12 (27.1) & 2.79E-12 ( 9.7) & 1.12E-11 (19.5) \\
\enddata
\tablecomments{Column (1): Name of the star [in AR~Lac, (f) refers to spectrum in flaring or 
 active stage]. Column (2) displays the spectra SWP 
 number. Column (3) and (4) represents exposure time and
 date, respectively, of the observation. Column (5) to (10)
 represent flux at Earth, in erg cm$^{-2}$ s$^{-1}$, and S/N
 in parenthesis [S/N is the signal to noise ratio, i. e.,
 the total flux in the line (minus corresponding continuum)
 divided by the square root of the quadratic summation of
 the errors associated with each point of the line].} 
\end{deluxetable}

\begin{deluxetable}{lllrrrrr}
\tablewidth{0pt}
\tabletypesize{\footnotesize}
\tablecaption{Electron Density Inferred from High Ionization Stages of
  Iron\label{tabne}}
\tablehead{
\colhead{} & \colhead{\ion{Fe}{22}} &
\multicolumn{2}{c}{\ion{Fe}{21}} & \ion{Fe}{20} &
\multicolumn{2}{c}{\ion{Fe}{19}}  &\\ 
\colhead{Name} & \colhead{$\frac{\lambda 114.41}{\lambda 117.17}$} &
\colhead{$\frac{\lambda 102.22}{\lambda 128.73}$} & \colhead{$
\frac{\lambda 142.16}{\lambda 102.22}$} & \colhead{$
\frac{\lambda 110.63}{\lambda 118.66+...}$} & 
\colhead{$\frac{\lambda 91.02}{\lambda 101.55+...}$} &
\colhead{$\frac{\lambda 91.02}{\lambda 108.37+...}$} & \colhead{Mean}}
\startdata
AY Cet  & \nodata & $12.7^{+0.4}_{-0.5}$ & \nodata & \nodata & \nodata
 & \nodata & 12.7 \\[4pt]
AR Psc  & $13.3^{+0.2}_{-0.4}$ & $12.7^{+0.2}_{-0.3}$ & \nodata & \nodata
 & \nodata & \nodata & 13.1 \\[4pt]
CC Eri  & $13.0^{+0.3}_{-0.5}$ & $13.0^{+0.2}_{-0.2}$ & \nodata
 & $13.6^{+0.2}_{-0.2}$ & \nodata & \nodata & 13.3 \\[4pt]
VY Ari  & \nodata & $12.4^{+0.2}_{-0.2}$ & $12.9^{+0.8}_{-0.7}$
 & $13.8^{+0.2}_{-0.2}$ & \nodata & \nodata & 13.4 \\[4pt]
$\epsilon$ Eri  & \nodata & $13.7^{+0.2}_{-0.2}$ & \nodata & \nodata & \nodata
 & $13.1^{+.09}_{-0.2}$ & 13.5 \\[4pt]
YY Gem  & $13.4^{+0.1}_{-0.2}$ & $13.0^{+0.2}_{-0.2}$ & \nodata & \nodata
 & \nodata & \nodata & 13.2 \\[4pt]
BF Lyn  & \nodata & $12.9^{+0.4}_{-0.7}$ & \nodata & \nodata & \nodata
 & \nodata & 12.9 \\[4pt]
DH Leo  & $12.6^{+0.4}_{-.09}$ & $11.9^{+0.4}_{-0.9}$ & \nodata & \nodata
 & \nodata & \nodata & 12.4 \\
\enddata
\tablecomments{Columns 5--7 corresponds to the line flux ratio of
  $\frac{\lambda 110.63}{\lambda 118.66+\lambda 121.83}$,
  $\frac{\lambda 91.02}{\lambda 101.55+\lambda 109.97+\lambda 111.70}$
  and $\frac{\lambda 91.02}{\lambda 108.37+\lambda 120.00}$
  respectively. Values expressed 
  represent log~$N_e$ in units of $cm^{-3}$, with error bars on the
  density obtained from 1-$\sigma$ errors on the flux measurements.
  AR Lac (q) and (f) corresponds to the ``quiescent'' and flaring observations
  of 1993+1997 and 2000 respectively. 
  } 
\end{deluxetable}

\begin{deluxetable}{llllllllllll}
\tabletypesize{\scriptsize}
\tabcolsep=2.5pt
\tablewidth{0pt}
\tablecaption{Emission Measure Distribution of stars in the sample
  (I)\label{tabemd}}
\tablehead{
\colhead{log (T)} & \multicolumn{11}{c}{log $\int N_e N_H dV$ (cm$^{-3}$)\tablenotemark{a}} \\ 
\colhead{($K$)} &  \colhead{$\beta$ Cet}  & \colhead{AY Cet} & \colhead{AR Psc} & \colhead{CC Eri} & \colhead{VY Ari} & \colhead{$\epsilon$ Eri} & \colhead{YY Gem} & \colhead{Procyon} & \colhead{BF Lyn} & \colhead{LQ Hya} & \colhead{DH Leo}}
\startdata
 4.0 & 51.85: & 52.20: & 52.30: & 51.20: & 52.60: & 50.30: & 50.70: & 51.40: & 51.40: & 51.60: & 52.00: \\
 4.1 & 51.80: & 52.15: & 52.00: & 50.90: & 52.30: & 50.10: & 50.65: & 51.30: & 51.30: & 51.30: & 51.80: \\
 4.2 & 51.75 & 52.10 & 51.90 & 50.80 & 52.00 & 49.90 & 50.60 & 51.10 & 51.20 & 51.10 & 51.60 \\
 4.3 & 51.65 & 52.05 & 51.70 & 50.60 & 51.80 & 49.75 & 50.50 & 50.80 & 51.10 & 50.80 & 51.40 \\
 4.4 & 51.50 & 52.00 & 51.60 & 50.40 & 51.50 & 49.60 & 50.40 & 50.70 & 50.90 & 50.60 & 51.20 \\
 4.5 & 51.30 & 51.95 & 51.40 & 50.20 & 51.30 & 49.50 & 50.35 & 50.60 & 50.70 & 50.40 & 50.90 \\
 4.6 & 51.20 & 51.90 & 51.15 & 49.80 & 50.90 & 49.35 & 50.30 & 50.55 & 50.40 & 50.20 & 50.50 \\
 4.7 & 51.00 & 51.80 & 51.10 & 49.60 & 50.60 & 49.15 & 50.00 & 50.10 & 50.25 & 50.00 & 50.25 \\
\enddata
\tablenotetext{a}{Emission measure, where $N_e$ 
and $N_H$ are electron and hydrogen densities, in $cm^{-3}$.
A colon indicates that the EMD value is uncertain because few lines 
occur in the temperature region.}
\end{deluxetable}

\begin{deluxetable}{llllllllllll}
\tabletypesize{\scriptsize}
\tabcolsep=2.5pt
\tablewidth{0pt}
\tablecaption{Emission Measure Distribution of stars in the sample (II)\label{tabemd2}}
\tablehead{
\colhead{log (T)} & \multicolumn{11}{c}{log $\int N_e N_H dV$
  (cm$^{-3}$)\tablenotemark{a}} \\
\colhead{($K$)} & \colhead{$\xi$ UMa} & \colhead{BH CVn} & \colhead{$\alpha$ Cen} & \colhead{$\sigma ^2$ CrB} & \colhead{V824 Ara} & \colhead{V478 Lyr} & \colhead{ER Vul} & \colhead{AR Lac\tablenotemark{b}} & \colhead{AR Lac\tablenotemark{c}} & \colhead{FK Aqr} & \colhead{BY Dra}}
\startdata
 4.0 & 51.30: & 52.60: & 49.40: & 52.40: & 52.20: & 51.50: & 52.50: & 53.01: & 53.10: & 50.40: & 51.40: \\
 4.1 & 50.90: & 52.40: & 49.30: & 52.20: & 52.00: & 51.30: & 52.20: & 52.70: & 52.70: & 50.30: & 51.10: \\
 4.2 & 50.70 & 52.30 & 49.20 & 52.10 & 51.80 & 51.10 & 52.00 & 52.40 & 52.50 & 50.20 & 50.80 \\
 4.3 & 50.50 & 52.20 & 49.10 & 51.80 & 51.60 & 50.90 & 51.80 & 52.10 & 52.30 & 50.10 & 50.70 \\
 4.4 & 50.40 & 52.00 & 49.05 & 51.50 & 51.30 & 50.60 & 51.60 & 51.80 & 52.10 & 50.00 & 50.60 \\
 4.5 & 50.35 & 51.80 & 49.00 & 51.30 & 51.20 & 50.40 & 51.40 & 51.60 & 51.90 & 49.90 & 50.40 \\
 4.6 & 50.30 & 51.60 & 48.90 & 51.25 & 51.10 & 50.20 & 51.00 & 51.40 & 51.70 & 49.80 & 50.20 \\
 4.7 & 50.15 & 51.50 & 48.70 & 51.10 & 51.00 & 50.00 & 50.85 & 51.25 & 51.65 & 49.60 & 49.90 \\
\enddata
\tablenotetext{a}{Emission Measure, where $N_e$ 
and $N_H$ are electron and hydrogen densities, in $cm^{-3}$. A colon
indicates that the EMD value is uncertain because few lines  
occur in the temperature region.}
\tablenotetext{b}{Only 1993 and 1997 observations are considered in
 this Emission Measure Distribution.} 
\tablenotetext{c}{Emission Measure Distribution from observations of the 2000 campaign.} 
\end{deluxetable}

\begin{deluxetable}{lrrr}
\tablewidth{0pt}
\tablecaption{Slope [d(log EM)/d(log T)] measured below the bump, and
bump height [$\Delta$(log EM)] over the slope for a given
temperature\label{slopes}}
\tablehead{\colhead{Star} & \colhead{Slope\tablenotemark{a}} & 
\colhead{Height} & \colhead{log T [K]}}
\startdata
$\beta$ Cet & \nodata &\nodata & \nodata \\
AY Cet      & 0.16$^c$ & 1.65 & 6.9 \\
AR Psc      & 0.67$^c$ & 0.32 & 7.0 \\
CC Eri      & 0.47$^b$ & 1.00 & 6.8 \\
VY Ari      & 1.12$^c$ & 0.87 & 6.9 \\
$\epsilon$ Eri & \nodata &\nodata & \nodata \\
YY Gem      & 0.58$^b$ & 0.97 & 6.8 \\
Procyon     & \nodata &\nodata & \nodata \\
BF Lyn      & 0.74$^d$ & 0.91 & 6.8 \\
LQ Hya      & 1.00$^c$ & 0.78 & 6.9 \\
DH Leo      & 0.86$^b$ & 0.88 & 6.8 \\
$\xi$ UMa   & \nodata & \nodata & \nodata \\
BH CVn      & \nodata & \nodata & \nodata \\
$\alpha$ Cen B & \nodata &\nodata & \nodata \\
$\sigma^2$ CrB & 1.70$^e$ & 0.78 & 6.7 \\
V824 Ara    & \nodata &\nodata & \nodata \\
V478 Lyr    & 1.11$^b$ & 1.17 & 6.8 \\
ER Vul      & 1.07$^b$ & 0.92 & 6.8 \\
AR Lac (q)  & 1.09$^c$ & 0.70 & 6.9 \\
AR Lac (f)  & 1.16$^c$ & 0.41 & 6.9 \\
FK Aqr      & 1.08$^f$ & 0.78 & 6.9 \\
BY Dra      & \nodata & \nodata & \nodata \\
\enddata
\tablenotetext{a}{Slopes calculated employing the values of log~T(K): 
  ($b$) 6.3, 6.4, 6.5, 7.1, 7.2, 7.3; 
  ($c$) 6.5, 6.6, 6.7, 7.1, 7.2, 7.3; 
  ($d$) 6.2, 6.3, 6.4, 6.9, 7.0, 7.1; 
  ($e$) 6.2, 6.3, 6.4, 7.1, 7.2, 7.3; 
  ($f$) 6.4, 6.5, 6.6, 7.1, 7.2, 7.3
}
\end{deluxetable}

\begin{deluxetable}{lrcrrrrrr}
  \tablewidth{0pt}
  \tablecolumns{5}
  \tablecaption{Bolometric and stellar fluxes of stars in the sample
  \tablenotemark{a}\label{tablelum}}
   \tablehead{\colhead{Star}  & \colhead{\bv} & 
        \colhead{b.c.\tablenotemark{b}}  &  \colhead{F$_{\rm bol}$} & 
        \colhead{F$_{\rm EUV}$}}
  \startdata
   $\beta$ Cet          & 1.03 & -0.41 &  5.547E-06 &  3.557E-12 \\
   AY Cet               & 0.89 & -0.28 &  2.154E-07 &  2.671E-12 \\
   AR Psc               & 0.83 & -0.22 &  4.022E-08 &  2.530E-12 \\
   CC Eri               & 1.39 & -0.90 &  1.883E-08 &  3.112E-12 \\
   VY Ari               & 0.96 & -0.34 &  6.718E-08 &  7.155E-12 \\
   UX Ari               & 0.88 & -0.27 &  8.197E-08 &  1.070E-11 \\
   $\epsilon$ Eri       & 0.88 & -0.27 &  1.022E-06 &  2.566E-12 \\
   V711 Tau             & 0.88 & -0.27 &  1.386E-07 &  1.202E-11 \\
\enddata
\tablenotetext{a}{Flux units are erg s$^{-1}$ cm$^{-2}$. Stars in the
sample displayed in Fig.~\ref{lumperiod}.} 
\tablenotetext{b}{Bolometric correction for the given \bv (see text).}
\tablenotetext{c}{EUVE flux corresponding to the 1993 and 1997
observations summed.} 

\end{deluxetable}

\end{document}